\documentclass[fleqn,usenatbib]{mnras}
\usepackage[T1]{fontenc}

\DeclareRobustCommand{\VAN}[3]{#2}
\let\VANthebibliography\thebibliography
\def\thebibliography{\DeclareRobustCommand{\VAN}[3]{##3}\VANthebibliography}

\usepackage{graphicx}	
\usepackage{amsmath}	
\usepackage{multirow}
\usepackage{amssymb}	
\usepackage{soul}
\usepackage[normalem]{ulem}

\newcommand{\bs}[1]{\boldsymbol{#1}}
\newcommand{\code}[1]{\texttt{#1}}

\title[A multitracer analysis for eBOSS using EFTofLSS]{A multitracer analysis for the eBOSS galaxy sample based on the effective field theory of large-scale structure}

\author[R. Zhao et al.]{
Ruiyang Zhao,$^{1,2,3}$
Xiaoyong Mu,$^{1,2}$
Rafaela Gsponer,$^{3}$
Jamie Donald-McCann,$^{3}$
\newauthor
\ Yonghao Feng,$^{1,2}$
Weibing Zhang,$^{1,2}$
Yuting Wang,$^{1,2}$
Gong-Bo Zhao,$^{1,2,4}$\thanks{E-mail: gbzhao@nao.cas.cn}
Kazuya Koyama,$^{3}$\thanks{E-mail: kazuya.koyama@port.ac.uk}
\newauthor
\ David Bacon$^{3}$
and Robert G. Crittenden$^{3}$
\\
$^{1}$National Astronomical Observatories, Chinese Academy of Sciences, Beijing, 100101, P.R.China\\
$^{2}$School of Astronomy and Space Science, University of Chinese Academy of Sciences, Beijing 100049, P.R.China\\
$^{3}$Institute of Cosmology \& Gravitation, University of Portsmouth, Dennis Sciama Building, Portsmouth, PO1 3FX, UK\\
$^{4}$Institute for Frontiers in Astronomy and Astrophysics, Beijing Normal University, Beijing 102206, P.R.China
}

\date{Accepted XXX. Received YYY; in original form ZZZ}

\pubyear{2023}

\begin{document}
\label{firstpage}
\pagerange{\pageref{firstpage}--\pageref{lastpage}}
\maketitle

\begin{abstract}
We perform a multitracer full-shape analysis in Fourier space based on the effective field theory of large-scale structure (EFTofLSS) using the complete Sloan Digital Sky Survey IV (SDSS-IV) extended Baryon Oscillation Spectroscopic Survey (eBOSS) DR16 luminous red galaxy (LRG) and emission line galaxy (ELG) samples. We study in detail the impact of the volume projection effect and different prior choices when doing the full-shape analysis based on the EFTofLSS model. We show that adopting a combination of Jeffreys prior and Gaussian prior can mitigate the volume effect and avoid exploring unphysical regions in the parameter space at the same time, which is crucial when jointly analysing the eBOSS LRG and ELG samples. We validate our pipeline using 1000 eBOSS EZmocks. By performing a multitracer analysis on mocks with comparable footprints, we find that cosmological constraints can be improved by $\sim10-35$ per cent depending on whether we assume zero stochastic terms in the cross power spectrum, which breaks the degeneracy and boosts the constraints on the standard deviation of matter density fluctuation $\sigma_8$. Combining with the Big Bang Nucleosynthesis (BBN) prior and fixing the spectral tilt $n_s$ to Planck value, our multitracer full-shape analysis measures $H_0=70.0\pm2.3~{\mathrm{km}}~{\mathrm{s}}^{-1}{\mathrm{Mpc}}^{-1}$, $\Omega_m=0.317^{+0.017}_{-0.021}$, $\sigma_8=0.787_{-0.062}^{+0.055}$ and $S_8=0.809_{-0.078}^{+0.064}$, consistent with the Planck~2018 results. In particular, the constraint on $\sigma_8$ is improved beyond that obtained from the single tracer analysis by $18$ per cent, or by $27$ per cent when assuming zero stochastic terms in the cross power spectrum.
\end{abstract}


\section{Introduction}
Massive spectroscopic galaxy surveys map the large-scale structure of the Universe through 3D clustering patterns of tracers including the baryonic acoustic oscillations \citep[BAO;][]{SDSS:2005xqv,2dFGRS:2005yhx} and redshift-space distortions \citep[RSD;][]{Kaiser:1987qv,Peacock:2001gs}. They are powerful probes of the cosmic expansion and the structure growth, respectively, and have been well measured from the two-point correlation function or power spectrum (PS) of the cosmic tracers \citep{BOSS:2016wmc,eBOSS:2020yzd}. These two-point statistics carry crucial information about cosmological parameters, therefore it is important to not only measure the two-point statistics accurately from observations, but also model them precisely in theory.

The power spectrum of galaxies is a convenient quantity for cosmological measurements, since Fourier modes on different scales are independent in the linear regime. The observational uncertainty of the PS consists of the cosmic variance (CV) and the shot noise, which dominate on large and small scales, respectively. If the level of shot noise is low, which is the case for stage-IV surveys such as Euclid \citep{Euclid:2011}, the Subaru Prime Focus Spectrograph \citep[PFS;][]{PFS:2014}, and the Dark Energy Spectroscopic Instrument \citep[DESI;][]{DESI:2016fyo}, CV's impact on parameters that are closely related to the amplitude of the power spectrum, such as the primordial non-Gaussianity parameter $f_\mathrm{NL}$ and RSD parameter can be suppressed \citep{Seljak:2008xr,McDonald:2008sh} when the PS of multiple tracers are jointly analysed because the dependence of the PS on $\delta_{\mathrm{m}}$, the overdensity field of matter, which is subject to the CV, can be cancelled out. This technique has been extensively studied in theory \citep{White:2008jy,Hector:2010,Bernstein:2011,Abramo:2012,Abramo:2013awa,Alarcon:2016bkr,Abramo:2021irg}. It has been applied to the GAMA and BOSS surveys by splitting galaxy samples according to colour and luminosity \citep{Blake:2013nif,Ross:2013vla} and the overlapping region of the Baryon Oscillation Spectroscopic Survey (BOSS) and the WiggleZ Dark Energy Survey \citep{Beutler:2015tla,Marin:2015ula}. Most recently, it was applied to the luminous red galaxies and emission line galaxies samples observed by the eBOSS survey \citep{Wang:2020tje,Zhao:2020tis}. We refer the readers to \citet{Wang:2020dtd} for a brief review on the application of multitracer method in the galaxy surveys.

The traditional way to extract cosmological information from the PS is based on the template fitting method \citep[e.g.][]{BOSS:2016psr}, namely, a given linear power spectrum is fed into the perturbation theory model to calculate the quasi-non-linear galaxy power spectrum in redshift space \citep[e.g.][]{Taruya:2010mx} under the assumption of galaxy biases \citep[e.g.][]{McDonald:2009}. We use the obtained non-linear galaxy power spectrum to fit data and get constraints on the Alcock--Paczynski (AP) and RSD parameters, with all bias parameters marginalized over. Then constraints on cosmological parameters can be derived from the AP and RSD parameters by assuming an underlying cosmological model. This template-based method is computationally efficient, since the evaluation of the template using perturbation theory, which is a time-consuming step, only needs to be performed once. However, the compression from the original data vectors of $P(k)$ to AP and RSD parameters may not be optimal \citep{Brieden:2021edu,Brieden:2021cfg,Chen:2021wdi,Maus:2023rtr}, therefore we may be subject to information loss when deriving cosmological parameters from the BAO and AP parameters.

A better approach is to derive cosmological parameters directly from the PS with more usable $k$ modes. Thanks to the development of the effective field theory of large-scale structure \citep[EFTofLSS, hereafter EFT;][]{Baumann:2010tm,Carrasco:2012cv,Ivanov:2022review}, and simulation-based emulators \citep{Zhai:2018plk,Nishimichi:2018etk} over the last decade, we are now able to obtain direct constraints on cosmological parameters from the PS, using many more modes in the quasi-non-linear regimes \citep{Ivanov:2019pdj,DAmico:2019fhj,Kobayashi:2021oud,Chen:2021wdi}. As we have discussed in the last paragraph, a full-shape analysis can extract information beyond the standard AP/RSD parameters, which motivates us to re-analyse the eBOSS LRG and ELG samples \citep{Zhao:2020tis} within the framework of EFTofLSS. In this work, we perform a Fourier-space multitracer analysis using the luminous red galaxies (LRGs) and emission line galaxies (ELGs) observed by the eBOSS survey in the framework of EFTofLSS.

The paper is structured as follows. The data sets and methodology used are presented in Section \ref{sec:data}, followed by a demonstration of our pipeline performed on mock catalogues in Section \ref{sec:mock}. The main result is presented in Section \ref{sec:results}, before discussions and conclusion in Section \ref{sec:conclusion}.

\section{Data and methodology}
\label{sec:data}
\subsection{Data set}
\begin{figure}
    \includegraphics[width=\columnwidth]{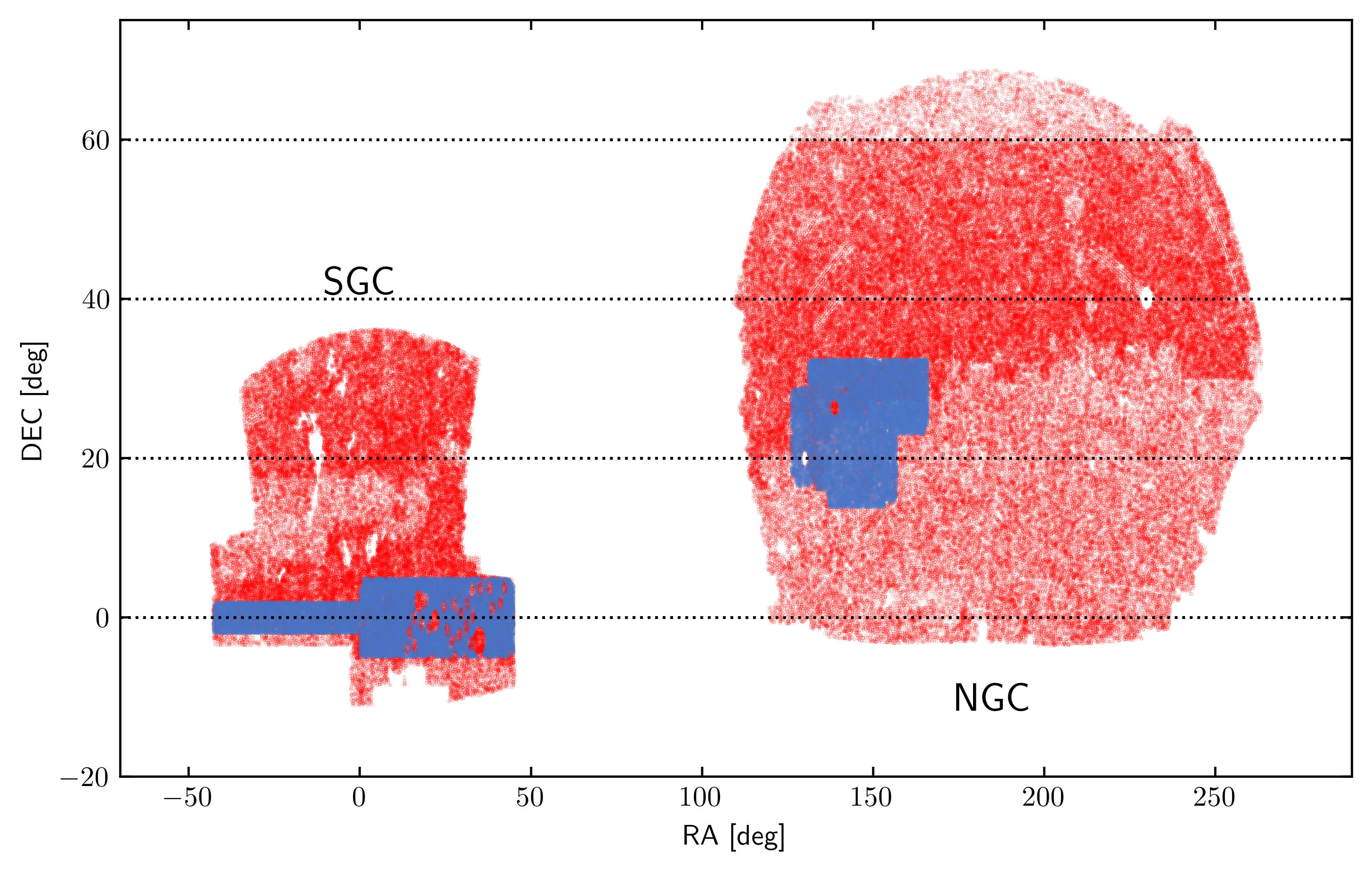}
    \caption{Footprint of LRG (red) and ELG (blue) samples in the North Galactic Cap (NGC) and South Galactic Cap (SGC) used in this analysis.}
    \label{fig:footprint}
\end{figure}

In the multitracer analysis performed in this study, we utilize the public eBOSS DR16 galaxy clustering catalogues of LRG and ELG samples\footnote{\url{https://data.sdss.org/sas/dr16/eboss/lss/catalogs/DR16/}}. The LRG catalogue used in this work is a combination of eBOSS LRGs and the BOSS DR12 LRGs observed at $z>0.6$, which is referred to as the ``CMASS+eBOSS LRG'' sample in the official eBOSS collaboration paper \citep[e.g.][]{Gil-Marin:2020bct,Bautista:2020ahg}. The detailed target selection algorithm and catalogue creation procedure are described in \citet{SDSS:2015gks}, \citet{Raichoor:2017nuz, Raichoor:2020vio}, and \citet{Ross:2020lqz}. We show the footprint of LRG and ELG samples in Fig.~\ref{fig:footprint} and list the relevant sample statistics in Table~\ref{tab:data_info}. The sky coverage of LRG samples is much larger than that of ELG, which means that the cosmological information is dominated by LRGs, as we will show in section~\ref{subsec:complete-mock}.

\begin{table*}
    \centering
    \caption{The relevant information for the auto- and cross-power spectrum of galaxy samples in the North Galactic Cap (NGC) and South Galactic Cap (SGC). $N_\text{data}$ is the number of objects in the galaxy clustering catalogue. $[z_\mathrm{min}, z_\mathrm{max}]$ is the redshift range utilized for measuring the power spectrum. $z_\mathrm{eff}$ is the effective redshift defined in equation~(\ref{eq:effective-redshift}), and $V$ is the comoving volume within that redshift range.}
    \begin{tabular}{c c c c c c c}
    \hline
    & LRG(NGC) & LRG(SGC) & ELG(NGC) & ELG(SGC) & Cross(NGC) & Cross(SGC)\\
    \hline\hline
    Effective area [$\mathrm{deg}^2$] & 6,934 & 2,560 & 370 & 358 & 370 & 358\\
    $N_\mathrm{data}$ & 255,741 & 121,717 & 83,769 & 89,967 & - & - \\
    $[z_\mathrm{min}, z_\mathrm{max}]$ & [0.6, 1.0] & [0.6, 1.0] & [0.6, 1.1] & [0.6, 1.1] & [0.6, 1.0] & [0.6, 1.0] \\
    $z_\mathrm{eff}$ & 0.696 & 0.705 & 0.849 & 0.841 & 0.763 & 0.774 \\
    $V\,[\mathrm{Gpc}h^{-1}]^3$ & 6.03 & 2.23 & 0.43 & 0.41 & - & - \\
    \hline
    \end{tabular}
    \label{tab:data_info}
\end{table*}

\subsection{Power spectrum measurement}
The anisotropic galaxy power spectrum of tracer $X$ and tracer $Y$ can be measured based on the Yamamoto estimator \citep{Yamamoto:2005dz}
\begin{equation}
    \hat{P}^{XY}_\ell(k) = \frac{2\ell + 1}{I^{XY}} \int \frac{d^2\hat{k}}{4\pi} F_\ell^X(-\bs{k})F_0^Y(\bs{k}) - P_{\ell,\mathrm{SN}}^{XY}(k)\text{,}
    \label{eq:Yamamoto-estimator}
\end{equation}
with
\begin{equation}
    F_\ell(\bs{k})=\int d^3x\,F(\bs{x}) \mathcal{L}_\ell(\hat{\bs{k}}\cdot\hat{\bs{\eta}})e^{-i\bs{k}\cdot\bs{x}}\text{,}
\end{equation}
where $\mathcal{L}_\ell$ represents the Legendre polynomial of the order of $\ell$ and $\hat{\bs{\eta}}$ is the line of sight (LOS). In this work we adopt the approximation taking $\hat{\bs{\eta}}$ to be $\hat{\bs{x}}$. The computational cost of $F_\ell(\bs{k})$ can be reduced by making use of the spherical harmonic addition theorem and factoring out the $k$-dependence in the Legendre polynomial \citep{Hand:2017irw}
\begin{equation}
    F_\ell(\bs{k}) = \frac{4\pi}{2\ell+1}\sum_{m=-\ell}^\ell Y_{\ell m}(\hat{\bs{k}})\int d^3x\,F(\bs{x})Y_{\ell m}^*(\hat{\bs{x}})e^{-i\bs{k}\cdot{\bs{x}}}\text{.}
\end{equation}
$F(\bs{x})$ is the FKP field \citep{Feldman:1993ky} defined as
\begin{equation}
    F(\bs{x}) = n_g(\bs{x}) - \alpha_s n_s(\bs{x})\text{,}
\end{equation}
where $n_g(\bs{x})$ and $n_s(\bs{x})$ represent the weighted number density of galaxy catalogue and random catalogue respectively, and $\alpha_s$ is the ratio of weighted numbers of data to random defined by
\begin{equation}
    \alpha_s = \frac{\sum_{i=1}^{N_g} w_{g,i}}{\sum_{i=1}^{N_s} w_{s,i}}\text{,}
\end{equation}
with the number of galaxies $N_g$ and randoms $N_s$, the total weight for each galaxy $w_{g,i}$ and random point $w_{s,i}$. The subtracted shot noise contribution $P_{\ell,\mathrm{SN}}^{XY}$ is only non-zero when $\ell=0$ and $X=Y$
\begin{equation}
    P_{\ell,\mathrm{SN}}^{XY} = \frac{\delta^K_{XY}\delta^K_{\ell0}}{I^{XY}} \left[ \sum_{i=1}^{N_g}w_{g,i}^2 + \alpha_s^2 \sum_{i=1}^{N_s} w_{s,i}^2 \right]\text{.}
\end{equation}

The normalization term $I^{XY}$ is evaluated by summing over mesh cells
\begin{equation}
    I^{XY} = \left[\alpha_s^Y\sum_i n_{g,i}^X n_{s,i}^Y + \alpha_s^X\sum_i n_{g,i}^Y n_{s,i}^X\right]\frac{dV}{2}\text{,}
    \label{eq:norm}
\end{equation}
where the weighted number densities are painted on the same mesh (of cell volume $dV$), using the Cloud In Cell (CIC) assignment scheme. We do not use the classic definition of normalization factor \citep[e.g.][]{BOSS:2016psr} evaluated by
\begin{equation}
    I^{XX} = \alpha_s \sum_{i=1}^{N_s} w_{s,i} \tilde{n}_{g,i}\text{,}
    \label{eq:classic-norm}
\end{equation}
where the summation is performed over the random catalogue and $\tilde{n}_{g,i}$ is the estimated weighted galaxy number density usually provided by the galaxy catalogue. As discussed in \citet{Zhao:2020bib}, due to the angular incompleteness, the classic normalization factor, which relies on the estimated $\tilde{n}_{g,i}$, may induce amplitude mismatching between the measured power spectrum with systematics and without systematics. Although the cosmological analysis remains unbiased with the same normalization for the measured window function \citep{deMattia:2019vdg}, the amplitude mismatching induces some subtleties when we subtract the radial integral constraint using mocks, as we will discuss in section~\ref{subsec:radial-integral-constraint}.

\begin{figure*}
    \includegraphics[width=0.95\linewidth]{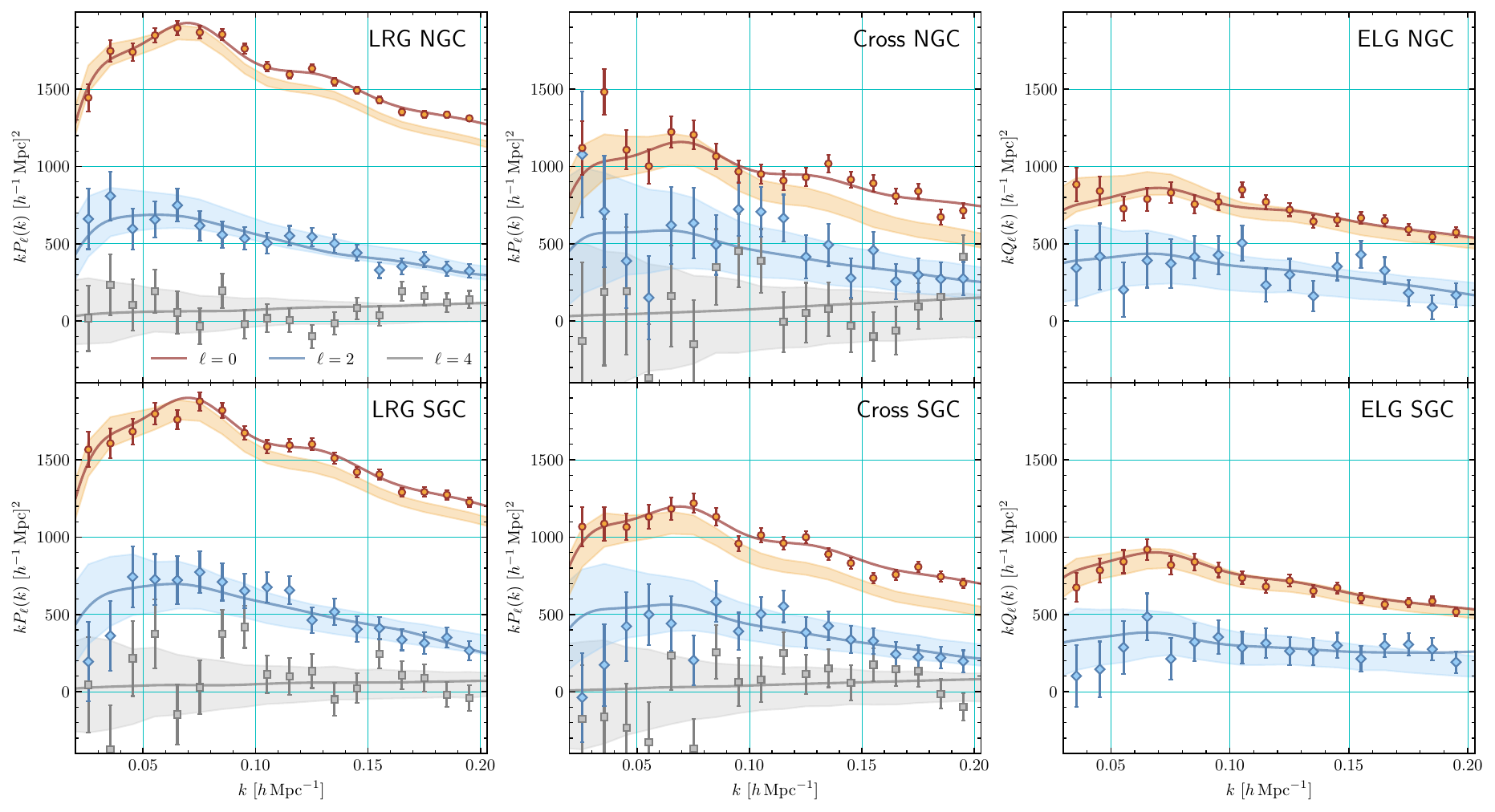}
    \caption{Power spectrum multipoles measured from eBOSS DR16 LRG and ELG data. Shadowed bands indicate the $1\sigma$ regions evaluated from 1000 EZmocks. Solid lines denote the best-fitting curves from our baseline analysis by performing a joint fit to all power spectrum measurements in both NGC and SGC. We show the chained power spectrum for ELG samples (see section~\ref{subsec:chained-pk}). Radial integral constraints have been subtracted in this plot (see section~\ref{subsec:radial-integral-constraint}).}
    \label{fig:data-fit}
\end{figure*}
We use the public code \code{pypower}\footnote{\url{https://github.com/cosmodesi/pypower}} to measure the power spectrum up to $k_\mathrm{max}=0.30\,h\mathrm{Mpc}^{-1}$, with $k$-bin $\Delta k=0.01\,h\mathrm{Mpc}^{-1}$. The FKP field is interpolated on a $512^3$ mesh with a box size of $5000\,h^{-1}\mathrm{Mpc}$ using the triangular shaped cloud (TSC) scheme. The $F_\ell(\bs{k})$ is computed with the fast Fourier transform (FFT) and we apply the TSC compensation window \citep{Jing:2004fq} and interlacing technique \citep{Sefusatti:2015aex} to correct the aliasing effect. In this paper, we use a fiducial $\Omega_m^\mathrm{fid}=0.307115$ to convert redshifts and celestial coordinates into comoving Cartesian coordinates. The fiducial cosmology dependence in the power spectrum measurement could be forward-modelled and corrected through equation (\ref{eq:AP-effect}).

Power spectrum measurements used in this work are displayed in Fig.~\ref{fig:data-fit}. The offset between data points and EZmocks (shadowed bands), which is evident at high $k$ for LRG and the cross power spectra, is caused by inappropriate normalization at the early stage of EZmock calibration and inaccurate modelling of cross-correlation between BOSS and eBOSS LRGs, LRGs and ELGs at small scales \citep[see][]{Zhao:2020bib}.

The effective redshifts of the measured power spectra are determined by summing over all weighted galaxy pairs with a separation between $25$ and $150\,h^{-1}\mathrm{Mpc}$,
\begin{equation}
    z_\mathrm{eff}^{XY} = \frac{\sum_{i,j}w_{g,i}^X w_{g,j}^Y (z_i^X + z_j^Y) / 2}{\sum_{i,j} w_{g,i}^X w_{g,j}^Y}\text{.}
    \label{eq:effective-redshift}
\end{equation}
The theory model will be evaluated at those effective redshifts; this assumes all galaxies are at the same redshift. In principle we could evaluate the weighted average of theory prediction according to the selection function \citep[see][]{Zhang:2021uyp}, but the difference was found to be small for BOSS samples, therefore we leave it for future work.

\subsection{Theory model}
According to the EFTofLSS model \citep{Perko:2016puo}, the galaxy density field in redshift space can be expanded perturbatively up to third order
\begin{equation}
    \delta_{g,r}=\delta_{g,r}^{(1)} + \delta_{g,r}^{(2)} + \delta_{g,r}^{(3)} + \delta_{g,r}^{(3,\mathrm{ct})} + \delta_{g,r}^{(\epsilon)}\text{,}
\end{equation}
where $\delta_{g,r}^{(1)}, \delta_{g,r}^{(2)}, \delta_{g,r}^{(3)}$ are related to the linear matter density field $\delta_m^{(1)}$ through the large-scale  bias expansion \citep[see][]{Desjacques:2016bnm} and $\delta_{g,r}^{(\epsilon)}$ represents the stochastic contribution from UV modes decoupled from large-scale observables. $\delta_{g,r}^{(3,\mathrm{ct})}$ consists of one ``speed of sound'' counter-term and two redshift-space counter-terms,  written as
\begin{equation}
    \delta_{g,r}^{(3,\mathrm{ct})} = (c_\mathrm{ct} k_M^{-2} + c_{r,1}\mu^2 k_R^{-2} + c_{r,2}\mu^4 k_R^{-2})k^2\delta_m^{(1)}\text{,}
\end{equation}
where $k_M^{-1}$ is the scale, typically the host halo scale, controlling the spatial derivative expansion, while $k_R^{-1}$ is the scale renormalizing the product of velocity operators arising from the real to redshift space transformation. Notice that $c_{r,2}$ is tracer independent and it is equal to the corresponding parameter for dark matter as discussed in \citet{Perko:2016puo}. However, since the effective redshift is different for LRG and ELG auto power spectra, we do not utilize this property when doing the fitting.

By correlating the density field of tracer $X$ and tracer $Y$, the $1$-loop power spectrum $P_{XY}(k,\mu)$ can be generalized to the cross correlation between two different tracers as follows
\begin{align}
    &\quad\  P_{XY}^\text{linear} + P_{XY}^\text{$1$-loop} + P_{XY}^\mathrm{ct} + P_{XY}^\mathrm{st}\\
    &= Z_1^X Z_1^Y P_{11}(k) + 2\int_{\bs{q}}Z_2^X Z_2^Y P_{11}(|\bs{k}-\bs{q}|)P_{11}(q) \nonumber\\
    &\phantom{=\qquad\qquad\qquad\ }+ 3P_{11}(k)\int_{\bs{q}}[Z_1^X Z_3^Y + Z_1^Y Z_3^X] P_{11}(q) \nonumber\\
    &+Z_1^X k^2 P_{11}(k)\left( c_\mathrm{ct}^X k^{-2}_{M,X} + c_{r,1}^X\mu^2 k_{R,X}^{-2} + c_{r,2}\mu^4 k^{-2}_{R,X}\right) + X\leftrightarrow Y\nonumber\\
    &+\frac{1}{2}(\bar{n}_X^{-1}+\bar{n}_Y^{-1})c_{\epsilon,0}^{XY} + \frac{1}{2}\left(\bar{n}_X^{-1} k_{M,X}^{-2} + \bar{n}_Y^{-1} k_{M,Y}^{-2}\right)k^2 \nonumber\\
    &\phantom{=\qquad\qquad\qquad\qquad\qquad}\times\left(c_{\epsilon,\text{mono}}^{XY} + c_{\epsilon,\text{quad}}^{XY}\mathcal{L}_2(\mu)\right)\text{,}
    \label{eq:power-spectrum}
\end{align}
with $P_{11}$ the linear matter power spectrum and $Z_n^X$ the $n$-th order redshift-space galaxy density kernels for tracer $X$, whose explicit forms are given by equation (A.2) in \citet{DAmico:2020kxu}. Equation (\ref{eq:power-spectrum}) restores the form of auto power spectrum if $X=Y$.

The stochastic term $P_{XY}^\mathrm{st}$ is parameterized as a constant (shot noise like) term and a $k^2$ term for monopole and quadrupole in the last line of equation~(\ref{eq:power-spectrum}), and it's normalised by the mean galaxy number density $\bar{n}$. Though this term in cross power spectrum is usually set to zero \citep[e.g.][]{Zhao:2020tis}, which assumes uncorrelated noise for different tracer populations, it may not be true in general cases. For example, as discussed in \citet{Mergulhao:2021kip}, features like the exclusion effect \citep[see][]{Baldauf:2013hka} have non-zero contribution to the stochastic term. Moreover, \citet{Alam:2019pwr} has reported a $3\sigma$ level detection of the $1$-halo galactic conformity with eBOSS data. They found that the occupation of LRGs and ELGs cannot be independent as assumed by the basic halo model, so they introduced an additional HOD parameter to capture the correlation between neighbouring galaxies belonging to the same halo. Given these considerations, the cross power spectrum model we use has three additional degrees of freedom and they cannot be expressed as the combination of stochastic terms appearing in the two auto power spectra.

We use a modified version of the \code{PyBird} code\footnote{\url{https://github.com/pierrexyz/pybird}} \citep{DAmico:2020kxu} to compute power spectrum multipoles. The linear power spectrum is calculated using the Boltzmann code \code{CLASS}\footnote{\url{https://github.com/lesgourg/class_public}} \citep{Blas:2011rf} and the loop integrals are evaluated using the FFTLog technique \citep{Simonovic:2017mhp}. To accurately model the BAO damping effect induced mainly by the long-wavelength displacement, the power spectrum is further IR-resummed following \citet{Senatore:2014via, Lewandowski:2015ziq, DAmico:2020kxu}. When comparing to data, since the measured power spectrum is a binned approximation to the continuous one, we take this effect into account by integrating the theory prediction within given $k$-bins (spherical shells).

The cross power spectrum model we use in this work is in general equivalent to that previously developed in \citet{Mergulhao:2021kip}, which has been extended to redshift-space recently \citep{Mergulhao:2023zso}. There are some minor differences mainly coming from the different parameterization between \code{PyBird} and \code{CLASS-PT} \citep{Chudaykin:2020aoj}. For instance, the definition of galaxy bias parameters are different [the relationship is given by equation~(1) in \citet{Simon:2022lde}] and the definition of EFT parameters in counter-terms and stochastic terms differs by a constant. In addition, since the mock test in section~\ref{sec:mock} does not display discernible bias, possibly attributed to the relatively higher redshift, we choose not to include other higher-in-$k$ or higher-in-$\mu$ counter-terms in this work.

\subsection{The Alcock--Paczynski effect}
As we assume a fiducial cosmology to convert redshifts and angular positions into comoving distances when measuring power spectrum, the difference between the fiducial and the underlying true cosmology produces a geometric distortion in $(k,\mu)$ space \citep{Alcock:1979mp}. The distorted power spectrum multipole reads \citep{Ballinger:1996cd}
\begin{align}
    P^\mathrm{AP}_\ell(k) &= \frac{2\ell+1}{2q_\perp^2 q_\parallel}\int_{-1}^1 P[k'(k,\mu), \mu'(k,\mu)]\mathcal{L}_\ell(\mu)\,d\mu \label{eq:AP-effect}\\
    k'(k,\mu) &= \frac{k}{q_\perp}\left[ 1 + \mu^2 (F^{-2} - 1) \right]^{1/2}\\
    \mu'(k,\mu) &= \frac{\mu}{F} \left[ 1 + \mu^2 (F^{-2} - 1) \right]^{-1/2} \text{,}
\end{align}
where the scaling parameters are given by
\begin{equation}
    q_\perp = \frac{D_M(z_\mathrm{eff})}{D_M^\mathrm{fid}(z_\mathrm{eff})}
    \quad
    q_\parallel = \frac{D_H(z_\mathrm{eff})}{D_H^\mathrm{fid}(z_\mathrm{eff})}
    \quad
    F = q_\parallel / q_\perp
    \text{,}
\end{equation}
with $D_M(z)$ the comoving angular diameter distance and $D_H(z) = c / H(z)$. The definition of scaling parameters does not depend on the BAO peak $r_d$ and Hubble constant because we are not fixing the linear power spectrum as a template and always working in the Hubble unit $h^{-1}\,\mathrm{Mpc}$ \citep[see discussions in][]{DAmico:2019fhj, eBOSS:2020yzd}. In practice, the 2D power spectrum $P(k',\mu')$ is constructed using multipoles up to $\ell=4$, since multipoles with $\ell>4$ have negligible contributions.

\subsection{Survey window}
The Yamamoto estimator we use measures the power spectrum convolved by the survey window \citep{Wilson:2015lup,Beutler:2018vpe}
\begin{equation}
    P_\ell^\text{win}(k) = \sum_{\ell'} \int W(k,k')_{\ell,\ell'}P_{\ell'}(k')\,dk'\text{.}
    \label{eq:window-convolve}
\end{equation}
The window matrix $W(k,k')_{\ell,\ell'}$ reads \citep{DAmico:2019fhj}
\begin{align}
    &W(k,k')_{\ell,\ell'} = \frac{2}{\pi} (-i)^\ell i^{\ell'} k'^2 \int s^2 j_\ell(ks)\mathcal{W}_{\ell,\ell'}(s)j_{\ell'}(k's)\,ds \\
    &\mathcal{W}_{\ell,\ell'}(s) = \sum_{\ell''} (2\ell + 1) 
        \begin{pmatrix}
        \ell & \ell' & \ell''\\
        0 & 0 & 0
        \end{pmatrix}^{2} \mathcal{W}_{\ell'}(s)\\
    &\mathcal{W}_{\ell}(s) \propto (2\ell + 1)\int\frac{d^2\hat{s}}{4\pi}\int d^3x\,\bar{n}_g(\bs{x}) \bar{n}_g(\bs{x} + \bs{s}) \mathcal{L}_\ell(\hat{\bs{s}} \cdot \hat{\bs{\eta}})\text{,}
\end{align}
where
$\begin{pmatrix}
    \ell & \ell' & \ell''\\
    0 & 0 & 0
\end{pmatrix}$ is the Wigner $3\text{-}j$ symbol, $\bar{n}_g$ is the selection function, $j_\ell$ represents the spherical Bessel function of order $\ell$ and $\mathcal{W}_{\ell}(s)$ denotes the multipole moment of the auto or cross-correlation function of survey window. The window matrix is evaluated using the FFTLog algorithm \citep{Hamilton:1999uv}, and in practice we perform the summation in equation~(\ref{eq:window-convolve}) up to $\ell'=4$.

\begin{figure*}
    \includegraphics[width=0.95\linewidth]{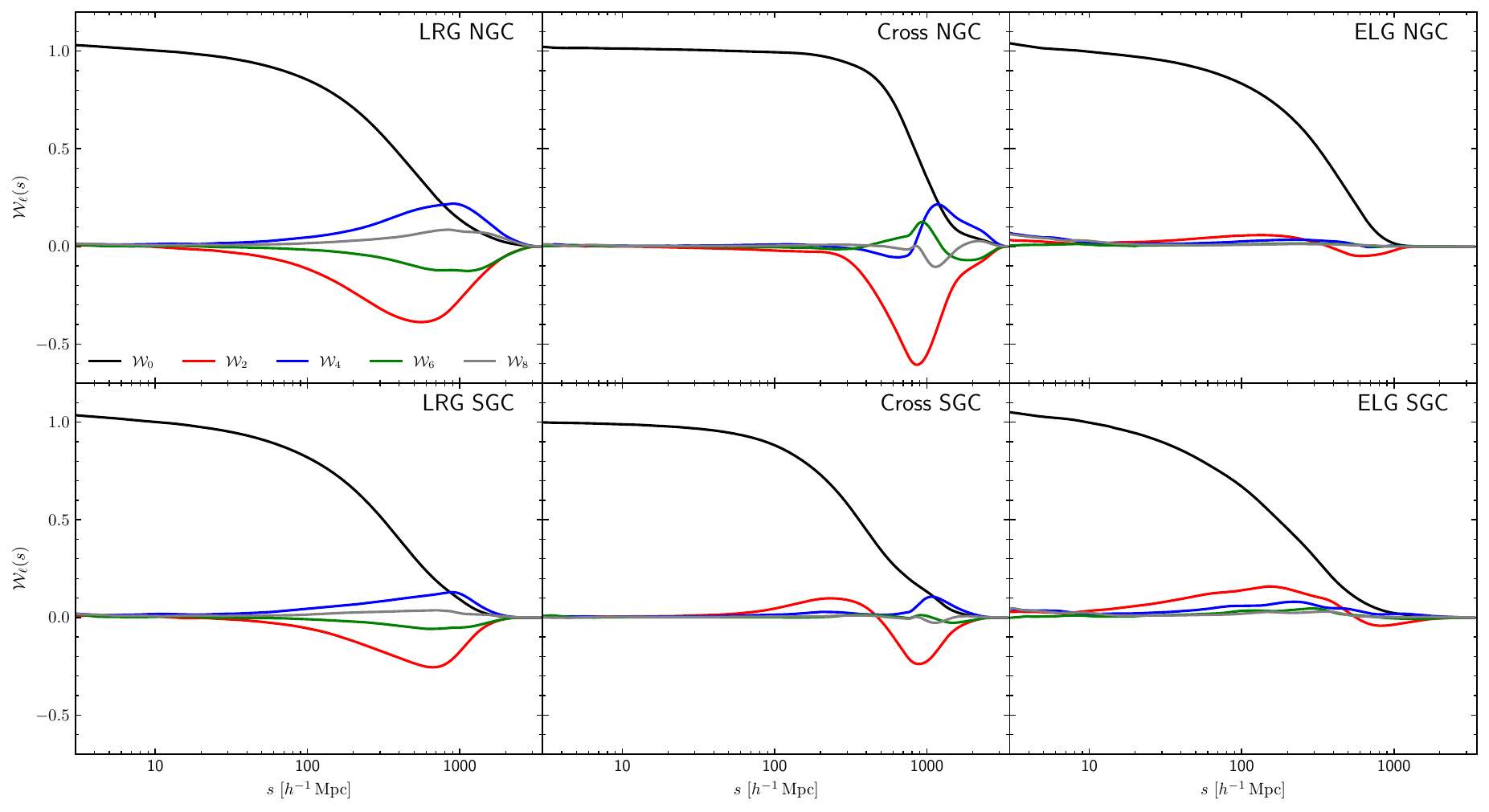}
    \caption{The auto (left and right panels) and cross (middle panel) window function multipoles of eBOSS samples in the North Galactic Cap (NGC) and South Galactic Cap (SGC). Points below $s=20\ h^{-1}\mathrm{Mpc}$ are smoothed by a third order Savitzky–Golay filter.}
    \label{fig:window}
\end{figure*}
Following \citet{Wilson:2015lup}, we adopt a pair-counting approach to estimate $\mathcal{W}_\ell(s)$
\begin{equation}
    \hat{\mathcal{W}}^{XY}_\ell(s) = \frac{2\ell+1}{I^{XY}}\alpha_s^X\alpha_s^Y \sum_{i,j}^{N_s}\frac{w_s^X(\bs{x}_i)w_s^Y(\bs{x}_j)}{4\pi s^3 \Delta\ln s} \mathcal{L}_\ell(\widehat{\bs{x}_j - \bs{x}_i} \cdot \hat{\bs{\eta}})\text{,}
\end{equation}
where $I^{XY}$ matches the normalization factor in equation~(\ref{eq:norm}) as suggested by \citet{deMattia:2019vdg}. The LOS $\hat{\bs{\eta}}$ is taken to be the mid-point of each galaxy pair instead of the end-point used in equation~(\ref{eq:Yamamoto-estimator}), however as studied in \citet{Beutler:2018vpe}, the error introduced by the inconsistent treatment of LOS only affects the very large scale, which is irrelevant in this work since we always limit our analysis to $k_\mathrm{min}\ge 0.02\,h\,\mathrm{Mpc}^{-1}$ modes. Figure~\ref{fig:window} shows the measured window function multipoles, note that the window functions of the ELG samples are anisotropic even at small scales due to the fine-grained veto masks \citep{deMattia:2020fkb}.

\subsection{The radial integral constraint}
\label{subsec:radial-integral-constraint}
The definition of galaxy overdensity field in the spectroscopic survey depends on our assumptions on the unclustered galaxy density field, namely the selection function, which is however difficult to determine without relying on the actual observed data. For example, we usually choose a selection function that nullifies the average density fluctuations within the entire survey volume, leading to the so-called global integral constraint \citep{Peacock:1991}. \citet{deMattia:2019vdg} proposed a model to correct this effect and they investigated an additional constraint, the radial integral constraint (RIC), because the radial selection function is also inferred indirectly from data, which is a dominant systematic in eBOSS ELG sample if not taken into account.

In this work, we adopt a simpler approach by estimating and subtracting the RIC effect using EZmocks. However, we caution that the direct subtraction from power spectra requires a consistent treatment of normalization, i.e. the expectation of the power spectrum estimator derived from equation~(\ref{eq:Yamamoto-estimator}) matches the unwindowed theoretical power spectrum at small scales (allowing for a same bias coefficient). The normalization defined in equation~(\ref{eq:norm}) approximately guarantees this property and is more accurate than the classic one defined in equation~(\ref{eq:classic-norm}) because it naturally accounts for the anisotropic selection function arising from the angular incompleteness. As shown in Fig.~\ref{fig:EZmock-pk}, we do not observe noticeable amplitude mismatching between power spectra measured from different EZmocks. We are aware that this approach relies on EZmocks and is an approximation. As demonstrated in \citet{Gsponer:2023}, RIC is a survey window-like effect, and the amount of correction depends on the power spectrum at large scales. However, since EZmocks match the clustering of data at large scales, we consider it a good approximation. After comparing with the results in \citet{Gsponer:2023}, we find that our independent pipelines yield consistent cosmological constraints within $\sim 0.5\sigma$ for the combined LRG and ELG auto power spectrum analysis.

\subsection{Chained power spectrum}
\label{subsec:chained-pk}
The eBOSS ELG samples suffered from unknown angular systematics due to the incomplete understanding of the survey selection function variation with imaging quality \citep{Raichoor:2020vio,Tamone:2020qrl,deMattia:2020fkb}. To mitigate this effect, we use the chained power spectrum multipole \citep{Hand:2017irw,Zhao:2020tis}
\begin{equation}
    Q_\ell = P_\ell - \frac{(2\ell + 1)\mathcal{L}_\ell(0)}{(2\ell + 5) \mathcal{L}_{\ell + 2}(0)} P_{\ell+2}\text{,}
\end{equation}
which is immune to any contaminant $S(k)$ coupling to the transverse mode, i.e.
\begin{equation}
    P^\mathrm{obs}(k,\mu) = P^\mathrm{true}(k,\mu) + S(k)\delta_D(\mu)\text{.}
\end{equation}
Although this is a simple model, it indeed captures the feature of angular systematics in the eBOSS ELG sample and was proven to work well in previous analysis \citep{Zhao:2020tis}. However, it is worth noting that the above formalism is not optimal due to the existence of the survey window and we refer readers to \citet{deMattia:2019vdg} and \citet{deMattia:2020fkb} for a more complete angular systematics mitigation method. Nevertheless, since the cosmological information is mainly dominated by the LRG sample, we find the shift caused by the remaining angular systematics in the chained power spectrum for the ELG sample is small as we will demonstrate in section~\ref{subsec:contaminated-mock}.

\subsection{Volume effect and analytical marginalization}
\label{sec:volume-effect-marg}
As discovered in previous studies \citep{DAmico:2022osl,Simon:2022lde,Chudaykin:2022nru,Carrilho:2022mon}, the EFTofLSS analysis usually suffers from the ``volume effect'', which is sometimes also called the ``projection effect''. This effect leads to biased cosmological constraints even when the data to fit is created by theory itself. Mathematically, the apparent bias arises because of the non-Gaussian full posterior, so the mean or the maximum point of the marginalized 1D or 2D posterior does not always coincide with the global maximum posterior. It typically happens when the data are not constraining enough, and this is why people found rescaling the covariance matrix by a small value could resolve the issue. In addition, different approaches used in the data analysis may suffer from different levels of the volume effect. For example, \citet{Maus:2023rtr} compared the direct model fitting to the traditional template based approach using the same EFT model. They found the degeneracies between cosmological parameters and nuisance parameters are more complicated in the former case, therefore it suffers from a stronger volume effect. The deeper implication of this finding still requires further study, but in principle, one can instead use the frequentist approach, the profile likelihood analysis \citep{Cousins:1994yw,Planck:2013nga,Herold:2021ksg}, to remove the volume effect, whose 1D credible intervals are by construction centred on the best-fitting parameters. However, this method is computationally expensive and requires high precision numerical code when searching for the extreme point. A recent paper \citep{Hadzhiyska:2023wae} points out that using the Jeffreys prior can partially cancel the volume effect and this cancellation is exact when nuisance parameters enter linearly in the model.

Following the notation and the derivation in \citet{DAmico:2020kxu} and \citet{Hadzhiyska:2023wae}, the EFT model reads
\begin{equation}
    P_\alpha = b_{G,i}P_{G,\alpha}^i + P_{NG,\alpha}\text{  ,}
\end{equation}
which is split into the Gaussian part $P_{G,\alpha}^i$ and the non-Gaussian part $P_{NG,\alpha}$. Model parameter $b_{G,i}$ is referred as the linear parameter or Gaussian parameter, since it appears linearly in the model. The index $\alpha$ runs over all possible $k$ and $\ell$ for both auto and cross power spectrum, and the repeated indices imply summation. The posterior can then be written as
\begin{align}
    -2\ln\mathcal{P} &= (P_\alpha - D_\alpha)C_{\alpha\beta}^{-1}(P_\beta - D_\beta)\nonumber\\
    &\qquad+ (b_{G,i} - \mu_{G,i})\sigma_{ij}^{-1}(b_{G,j} - \mu_{G,j}) - 2\ln\Pi\text{,}
    \label{eq:full-posterior}
\end{align}
where $D_\alpha$ is the data vector, $C_{\alpha\beta}$ is the data covariance matrix, $\bs{b}_G$ has a Gaussian prior $\bs{b}_G \sim \mathcal{N}(\bs{\mu}_G, \sigma)$ and $\Pi$ includes the normalization coefficient and the prior of non-Gaussian parameters. By doing the Gaussian integral, the marginalized posterior of non-linear parameters is expressed as
\begin{equation}
    -2\ln\mathcal{P}_\text{marg} = \chi^2_* + \chi^2_{*,G} -2\ln\Pi + \ln\det\left(\frac{F_2}{2\pi}\right)\text{,}
    \label{eq:marg}
\end{equation}
with
\begin{align}
    &b_{G,i}^* = F_{2,ij}^{-1}F_{1,j}\label{eq:bG_best}\\
    &P^*_\alpha = b_{G,i}^* P_{G,\alpha}^i + P_{NG,\alpha}\\
    &\chi^2_* = (P^*_\alpha - D_\alpha) C_{\alpha\beta}^{-1} (P^*_\beta - D_\beta)\\
    &\chi^2_{*,G} = (b_{G,i}^* - \mu_{G,i})\sigma_{ij}^{-1}(b_{G,j}^* - \mu_{G,j})\\
    &F_{2,ij} = P_{G,\alpha}^i C_{\alpha\beta}^{-1} P_{G,\beta}^j + \sigma_{ij}^{-1}\\
    &F_{1,i} = -P_{G,\alpha}^i C_{\alpha\beta}^{-1}(P_{NG,\beta} - D_\beta)  +\sigma_{ij}^{-1}\mu_{G,j}\text{  .}
\end{align}
$b_{G,i}^*$ and $F_{2,ij}$ represent the best-fitting value of linear parameters and the Fisher matrix, respectively, when fixing non-linear parameters. When data are noisy, $\chi^2_*\sim\mathrm{const.}$ therefore the last term in equation~(\ref{eq:marg}) dominates the posterior (neglecting the Gaussian prior $\chi^2_{*,G}$). However, since $F_{2}$ is the Fisher matrix and does not depend on the actual data, the constraining power provided by this term does not come from data but from our specific choice of parameterization, which introduces the volume effect. This can be exactly cancelled out by adding the Jeffreys prior\footnote{This is only true when marginalising over linear parameters \citep[see][]{Hadzhiyska:2023wae}.} (differs by a constant)
\begin{equation}
    -2\ln\Pi_\mathrm{Jeff.} = -\ln\det \left(\frac{F_{2}}{2\pi}\right)+\mathrm{constant}\text{.}
\end{equation}

We are aware that although the marginalized joint posterior of non-linear parameters does not suffer from the volume effect caused by linear parameters with this method, the 1D posterior of cosmological parameters may still be biased due to the remaining non-trivial degeneracies between non-linear parameters. In principle, we can do analytical marginalization and remove the volume effect caused by all nuisance parameters based on Laplace approximation \citep{Hadzhiyska:2023wae}. We do not investigate this approach in this paper and leave it for future work. When summarizing cosmological constraints, we choose to report the $68$ per cent credible interval as well as the best-fitting point, i.e. the maximum \textit{a posteriori} (MAP)\footnote{Here, we minimize the marginalized posterior, but it is equivalent to minimizing the full posterior because we have included the Jeffreys prior.}, which is, by definition, not sensitive to the volume effect \citep{DAmico:2022osl,Simon:2022lde} as a self-diagnostic way of checking the remaining volume effect.

\subsection{Parameter estimation}
\label{subsec:parameter-estimation}
In our analysis, the full EFT model includes the following nuisance parameters:
\begin{equation}
    \{ b_1^\alpha, b_2^\alpha, b_3^\alpha, b_4^\alpha \} \times \{ c_{ct}^\alpha, c_{r,1}^\alpha, c_{r, 2}^\alpha \} \times \{ c_{\epsilon, 0}^\beta, c_{\epsilon, \mathrm{mono}}^\beta, c_{\epsilon, \mathrm{quad}}^\beta \}\text{,}
\end{equation}
where for both NGC and SGC, the index $\alpha$ runs over LRG and ELG, $\beta$ runs over LRG, ELG and Cross. We allow those parameters to be different in different skycuts due to the different target selection criteria, giving 46 free parameters in total. $b_1$ is the linear bias and $b_2,b_3,b_4$ are higher-order bias parameters \citep{Perko:2016puo}, $c_{ct},c_{r,1},c_{r,2}$ are counter-terms and $c_{\epsilon,0}, c_{\epsilon, \mathrm{mono}}, c_{\epsilon, \mathrm{quad}}$ are stochastic terms. As discussed in \citet{DAmico:2019fhj}, $b_2$ and $b_4$ are highly degenerate, thus we define and sample on the following new parameters
\begin{equation}
    c_2 = \frac{1}{\sqrt{2}}(b_2 + b_4)\qquad c_4 = \frac{1}{\sqrt{2}} (b_2 - b_4)\text{.}
\end{equation}
However, the signal-to-noise ratio of present-day data does not allow us to measure $c_4$ and $c_{\epsilon,\mathrm{mono}}$, so we set both of these to zero. The non-linear and renormalization scale are chosen to be $k_M=0.7\ h\,\mathrm{Mpc}^{-1}$ and $k_{R}=0.25\ h\,\mathrm{Mpc}^{-1}$ \citep{Simon:2022lde}, and the mean galaxy number densities (in units of $[\mathrm{Mpc}/h]^{-3}$) are estimated as
\begin{align}
    \text{NGC:  }&\bar{n}_\mathrm{LRG}=4.5\times 10^{-5},\quad \bar{n}_\mathrm{ELG}=2.3\times 10^{-4}; \\
    \text{SGC:  }&\bar{n}_\mathrm{LRG}=5.9\times 10^{-5},\quad
    \bar{n}_\mathrm{ELG}=2.5\times 10^{-4}\text{.}
\end{align}
We caution that the cross power spectrum shares the same set of nuisance parameters with LRG and ELG auto power spectrum. This is an approximation because they are computed at different effective redshifts and we neglect the redshift evolution of EFT parameters. Since the discrepancy in redshift is at $\sim0.07$ level, this approximation proves to work well as we will show later in section~\ref{subsec:complete-mock}.

In subsequent sections, the LRG power spectrum, ELG power spectrum and the cross power spectrum between LRGs and ELGs are abbreviated as L, E and X, respectively, when they are used as indices, e.g. $b_1^\mathrm{LN}$ denotes the linear bias parameter of the LRG sample in the NGC. Different data sets are abbreviated as follows
\begin{itemize}
    \item LRG: LRG power spectrum
    \item ELG: ELG power spectrum
    \item LpE: LRG + ELG power spectrum
    \item LEX: LRG+ELG+Cross power spectrum.
\end{itemize}
All data sets include both NGC and SGC unless otherwise stated.

\begin{figure*}
    \includegraphics[width=0.95\linewidth]{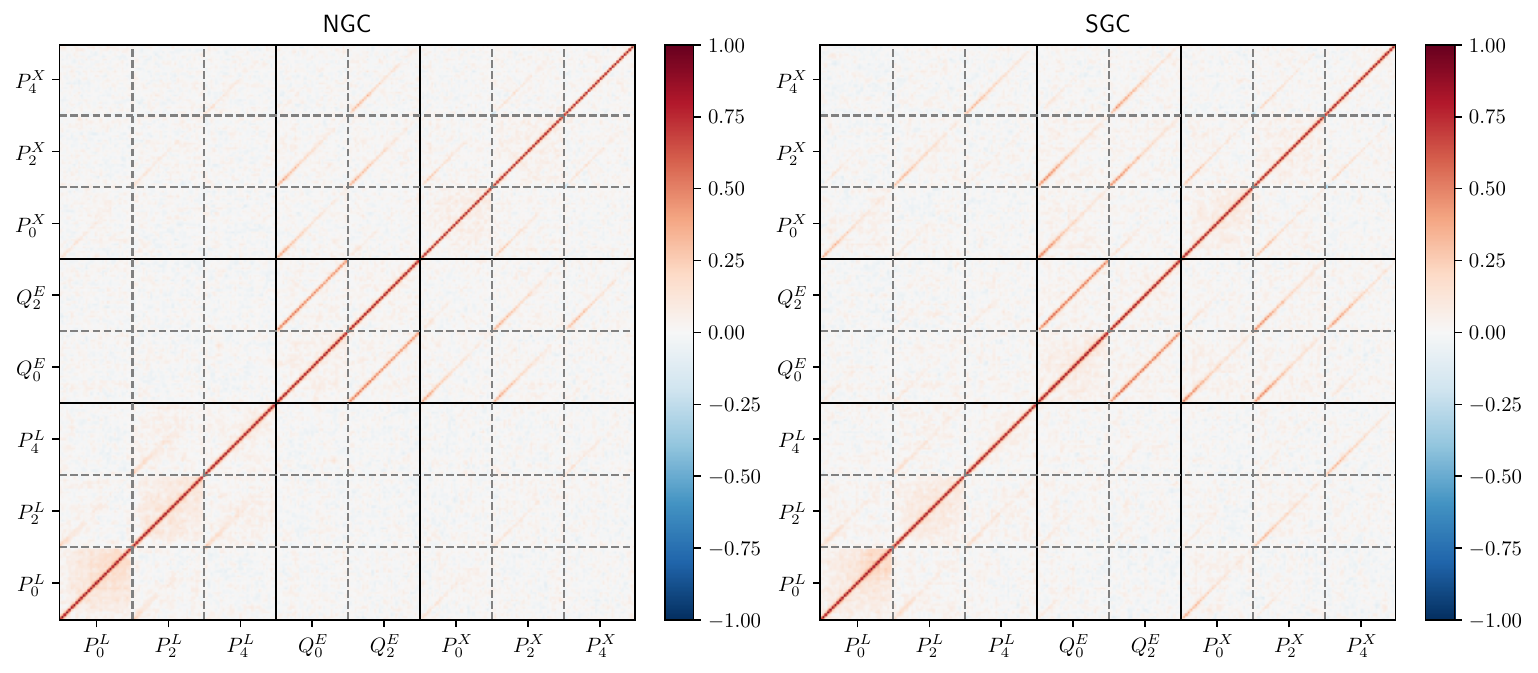}
    \caption{The correlation matrix of power spectrum $P_\ell$ and chained power spectrum $Q_\ell$ between LRG (L), ELG (E) and Cross (X). The left panel shows the North Galactic Cap (NGC) and the right panel shows the South Galactic Cap (SGC). In this plot, we use modes between $k_\mathrm{min}=0$ and $k_\mathrm{max}=0.30\ h\,\mathrm{Mpc}^{-1}$. LRG power spectrum is less correlated with ELG power spectrum and the cross power spectrum because LRGs cover a much wider sky area than ELGs.}
    \label{fig:corr_full}
\end{figure*}
We use a Gaussian likelihood for parameter estimation. The NGC and SGC data sets are treated independently, but we use a large covariance matrix computed from 1000 contaminated EZmocks mentioned in section~\ref{subsec:contaminated-mock} to capture the correlation between different samples (see Fig.~\ref{fig:corr_full}). In our analysis, the full data vector is given by
\begin{equation}
    \bs{D} = \{ P_0^L, P_2^L, P_4^L, Q_0^E, Q_2^E, P_0^X, P_2^X, P_4^X \}\text{,}
\end{equation}
with $k_\mathrm{min}=0.02\ h\,\mathrm{Mpc}^{-1}$ for LRG and Cross, while $k_\mathrm{min}=0.03\ h\,\mathrm{Mpc}^{-1}$ for ELG, and the same $k_\mathrm{max}=0.20\ h\,\mathrm{Mpc}^{-1}$ for all three samples. As the covariance matrix is estimated from a finite number of mocks, the inverse covariance matrix should be further rescaled by the Hartlap factor \citep{Hartlap:2006kj}
\begin{equation}
    \tilde{\bs{C}}^{-1} = \frac{N_r - N_d - 2}{N_r - 1} \bs{C}^{-1}\text{.}
\end{equation}
In our analysis, the number of mock realizations $N_r=1000$ and the number of data points $N_d=147$, leads to a $0.85$ level correction. As discussed in section~\ref{sec:volume-effect-marg}, we perform analytical marginalization over the following linear parameters
\begin{equation}
    \bs{b}_G = \{ b_3^\alpha, c_{ct}^\alpha, c_{r,1}^\alpha, c_{r,2}^\alpha, c_{\epsilon, 0}^\beta,  c_{\epsilon, \mathrm{quad}}^\beta \}\text{,}
    \label{eq:linear-parameters}
\end{equation}
using equation~(\ref{eq:marg}), dropping the last term related to the volume effect. Finally, we directly sample over the following $12$-dimensional parameter space, assuming the base flat $\Lambda$CDM model:
\begin{equation}
    \bs{\Omega} = \{ \omega_b, \omega_{cdm}, H_0, \ln (10^{10}A_s) \} \times \{ b_1^\alpha, c_2^\alpha\}\text{.}
\end{equation}

\begin{table}
    \centering
    \caption{The prior information for varied cosmological parameters.  $\mathcal{U}(\mathrm{min}, \mathrm{max})$ and $\mathcal{N}(\mu, \sigma^2)$ denote the uniform and Gaussian prior respectively.}
    \begin{tabular}{cccc}
    \hline
    $\omega_b$ & $\omega_{cdm}$ & $H_0$ & $\ln(10^{10}A_s)$ \\
    \hline\hline
    $\mathcal{N}(0.02268, 0.00038^2)$ & $\mathcal{U}(0.03, 0.7)$ & $\mathcal{U}(40, 100)$ & $\mathcal{U}(0.1, 10)$ \\
    \hline
    \end{tabular}
    \label{tab:cosmo-prior}
\end{table}
The prior information for cosmological parameters is listed in Table~\ref{tab:cosmo-prior}. We apply a Gaussian prior on the physical baryon density $\omega_b=0.02268\pm0.00038$ as used in \citet{Ivanov:2019pdj}, which is inferred from the standard BBN analysis \citep{Aver:2015iza,Cooke:2017cwo} assuming $N_\mathrm{eff}=3.046$. We fix the spectral index\footnote{LRG data alone is able to constrain $n_s$, but the constraint is notably weaker, by two orders of magnitude, than Planck~2018 result.} to the Planck~2018 value $n_s=0.965$ \citep{Planck:2018vyg} and assume a single massive neutrino with mass $m_\nu=0.06\,\mathrm{eV}$\footnote{We only take the neutrino effect into account at leading order through the linear power spectrum $P_{11}(k)$, and we use the CDM~+~baryons power spectrum as suggested by \citet{Vagnozzi:2018pwo}}. While in section~\ref{sec:mock}, we fix $\omega_b, n_s$ and $\sum m_\nu$ to the fiducial values used in the mocks.

\begin{table*}
    \centering
    \caption{Same as Table~\ref{tab:cosmo-prior}, but for EFT parameters. Index $\alpha$ runs over LRG and ELG, $\beta$ runs over LRG, ELG and Cross, and for both NGC and SGC. $\Pi_A$ is the uninformative prior. $\Pi_B$ prior assumes a classic Gaussian prior as used in \citet{Simon:2022lde}, which enforces the perturbative condition. Both $\Pi_A$ and $\Pi_B$ are combined with the Jeffreys prior in order to mitigate the volume effect.}
    \begin{tabular}{c cc cccccc}
    \hline
    & $b_1^\alpha$ & $c_2^\alpha$ & $b_3^\alpha$ & $c_{ct}^\alpha$ & $c_{r,1}^\alpha$ & $c_{r,2}^\alpha$ & $c_{\epsilon, 0}^\beta$ & $c_{\epsilon, \mathrm{quad}}^\beta$ \\
    \hline\hline
    $\Pi_A$ & $\mathcal{U}(0, 4)$ & $\mathcal{U}(-100, 100)$ & -- & -- & -- & -- & -- & -- \\
    \hline
    $\Pi_B$ & $\mathcal{U}(0, 4)$ & $\mathcal{U}(-100, 100)$ & $\mathcal{N}(0, 4^2)$ & $\mathcal{N}(0, 2^2)$ & $\mathcal{N}(0, 4^2)$ & $\mathcal{N}(0, 4^2)$ & $\mathcal{N}(0, 2^2)$ & $\mathcal{N}(0, 2^2)$ \\
    \hline
    \end{tabular}
    \label{tab:nuisance-prior}
\end{table*}
The prior information for all varied EFT parameters is displayed in Table~\ref{tab:nuisance-prior}. $\Pi_A$ denotes the uninformative prior, we do not make any assumptions about linear parameters with an infinitely wide prior, which avoids choosing hyperparameters in the prior (also the $k_M$, $k_R$ and $\bar{n}$) and thus avoids the prior weight effect \citep{Simon:2022lde}. However, this is a very conservative choice and might include some unphysical regions in the parameter space, e.g. regions with extremely large counter-terms, where the perturbation theory breaks down. However, defining a prior which satisfies the perturbative condition always involves some ambiguities. Though we know the value of those parameters are $\sim\mathcal{O}(1)$ by construction, this is a rough estimate. When the data are strong enough, the choice of prior does not matter, but this is not the case for eBOSS data when considering this kind of EFT parameterization. We simply adopt a classic Gaussian prior as used in \citet{Simon:2022lde}, which is a common choice in many EFT analyses, to ensure that the theory is perturbative. Ideally, we can analyse the degeneracy and set some parameters to zero as we did for $c_4$, or select relevant parameters based on the Bayesian evidence, but those are beyond the scope of this paper and we refer the readers to \citet{donald-mccann:2023} for a detailed discussion on sub-model selection. Therefore, in addition to the $\Pi_A$ prior, we also report results analysed with the perturbative prior ($\Pi_B$), i.e. the classic Gaussian prior\footnote{We use a wider prior for $b_3$, as we find that the EZmock analysis prefers a high value of $b_3$. We also include $c_{r,2}$, which is relevant for hexadecapole.} \citep{Simon:2022lde} combined with the Jeffreys prior.

We perform the Bayesian analysis using the \code{Cobaya}\footnote{\url{https://github.com/CobayaSampler/cobaya}} code \citep{Torrado:2020dgo} with the default adaptive fast-slow MCMC sampler \citep{Lewis:2002ah,Lewis:2013hha}. The sampling is further accelerated with the fast-dragging technique \citep{Neal:2005math} taking advantage of the large speed hierarchy between fast EFT parameters and slow cosmological parameters. All chains are checked to reach the Gelman-Rubin criterion \citep{Gelman:1992zz} with $R-1<0.01$ to ensure the convergence. We use \code{Py-BOBYQA} \citep{Cartis:2018code} to find the maximum \textit{a posteriori} (MAP) and use the \code{GetDist}\footnote{\url{https://github.com/cmbant/getdist}} \citep{Lewis:2019xzd} package to plot the marginalized posterior.

\section{Tests on mock catalogues}
\label{sec:mock}

\subsection{Synthetic mock}
\label{subsec:synthetic-mock}
\begin{figure}
    \includegraphics[width=\columnwidth]{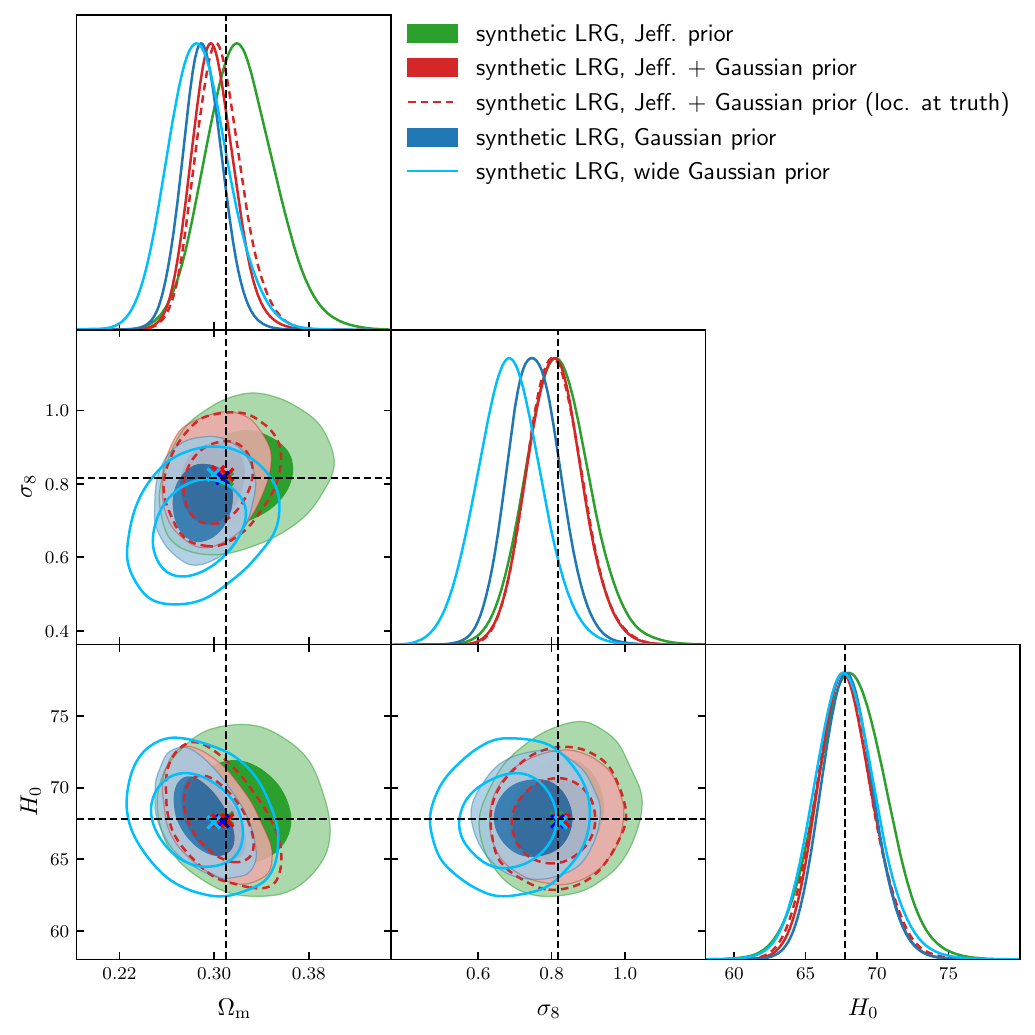}
    \caption{Comparison of cosmological posteriors derived from different prior choices using the eBOSS LRG-like synthetic data. The black dashed lines represent the fiducial cosmology and the cross markers denote the MAPs. The filled green and filled red contours correspond to $\Pi_A$ and $\Pi_B$ prior respectively (see Table~\ref{tab:nuisance-prior}). The filled blue contour denotes the classic Gaussian prior (no Jeffreys prior) widely used in the literature.}
    \label{fig:synthetic-data}
\end{figure}
In order to test the volume projection effect, we create one synthetic mock using our pipeline code. The cosmological parameters and EFT parameters are determined by fitting to the mean of 1000 \textit{complete} eBOSS LRG EZmocks (see section~\ref{subsec:complete-mock}) for both NGC and SGC. The fitted nuisance parameters generally fall within $1\sigma$ of the $\Pi_B$ prior given in Table~\ref{tab:nuisance-prior}. Then we fit to the synthetic data with different prior choices and present the corresponding cosmological constraints in Fig.~\ref{fig:synthetic-data}.

The filled green contour and the filled red contour in the figure correspond to the Jeffreys prior ($\Pi_A$) and the Jeffreys prior + Gaussian prior ($\Pi_B$), respectively. The blue contour represents the classic Gaussian prior without including the Jeffreys prior. After applying the Jeffreys prior, the mean of $\sigma_8$ is less biased, and by comparing to the case with a wide Gaussian prior $\sim\mathcal{N}(0, 100^2)$, this mitigation does not simply result from the inflated contour. This means our method can indeed mitigate the volume effect. The $\Pi_B$ prior also works well, but it suffers from a small prior weight effect by comparing to the red dashed line, where the Gaussian prior centres at the truth value of nuisance parameters. The mean of $\Omega_m$ is not unbiased for neither $\Pi_A$ nor $\Pi_B$, indicating there are remaining volume effects caused by the nontrivial degeneracy between non-linear parameters, but they are still consistent with the truth within $1\sigma$. Cross markers in the figure denote the MAPs and they are close to the truth in all cases.

The difference between $\Pi_A$ and $\Pi_B$ provides a rough estimate of how many unphysical regions in the parameter space are excluded if we respect the perturbative condition. In our specific case, regions favouring a high $\Omega_m$ value are excluded. In extreme situations, such as when analysing the BOSS z1 SGC multipole measurements with the Jeffreys prior, \citet{donald-mccann:2023} observed an unusual double-peak feature in the marginalized posterior where EFT parameters related to one of these modes favour more extreme values. Applying the perturbative condition can remove it\footnote{Differently from the Gaussian prior we use in this work, \citet{donald-mccann:2023} uses a different perturbative condition enforcing the loop and counter terms to be smaller than the linear part. We refer the readers to \citet{Braganca:2023pcp} for an alternative form of the perturbative prior.}. This is a limitation of the uninformative Jeffreys prior making our constraints overconservative due to the inclusion of some unphysical regions.

\begin{figure*}
    \includegraphics[width=0.95\linewidth]{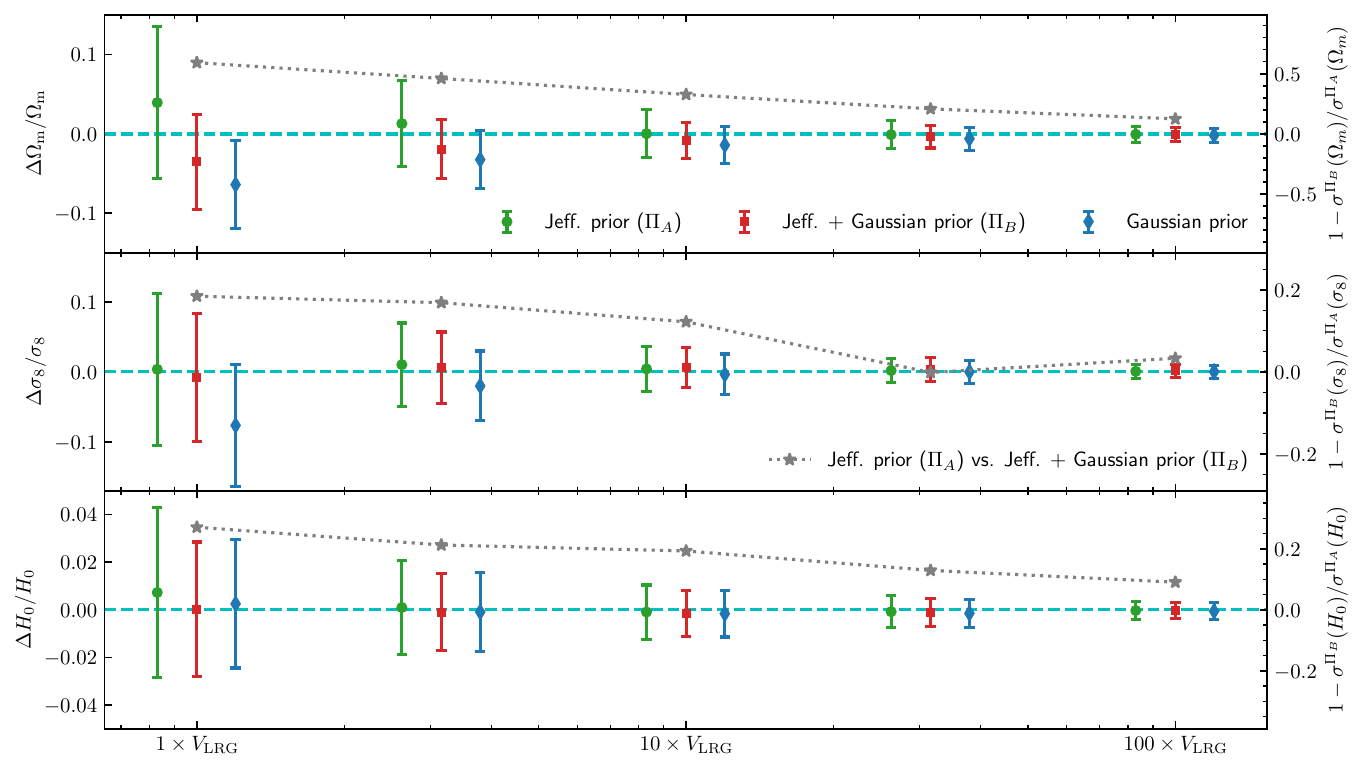}
    \caption{The mean and standard deviation of cosmological parameters extracted from the 1D marginalized posterior analysed using eBOSS LRG-like synthetic data, displayed as a function of survey volume. For better visibility, values are compared to and normalised by the truth. The grey dotted line shows the fractional difference of cosmological constraints between using the Jeffreys prior and Jeffreys prior + Gaussian prior as indicated on the right hand side of the plot and show the additional constraining power due to the Gaussian prior.}
    \label{fig:synthetic-data-varying-V}
\end{figure*}
We then study the impact of data quality on cosmological constraints by rescaling the covariance matrix. Our results are displayed in Fig.~\ref{fig:synthetic-data-varying-V}. As expected, when data are strong enough, different prior choices are all unbiased. However, when comparing the error bars of cosmological parameters using the Jeffreys prior (green) and the Jeffreys prior + Gaussian prior (red), we find the Gaussian prior enforcing the perturbative condition provides non-negligible information even when we have a survey volume that is $100$ times that of eBOSS. The detailed study of choosing reasonable hyperparameters in the Gaussian prior is beyond the scope of this paper, so we choose to report both the conservative results using the $\Pi_A$ prior and the more aggressive results using the $\Pi_B$ prior.

\subsection{Complete EZmock}
\label{subsec:complete-mock}
\begin{figure*}
    \includegraphics[width=0.95\linewidth]{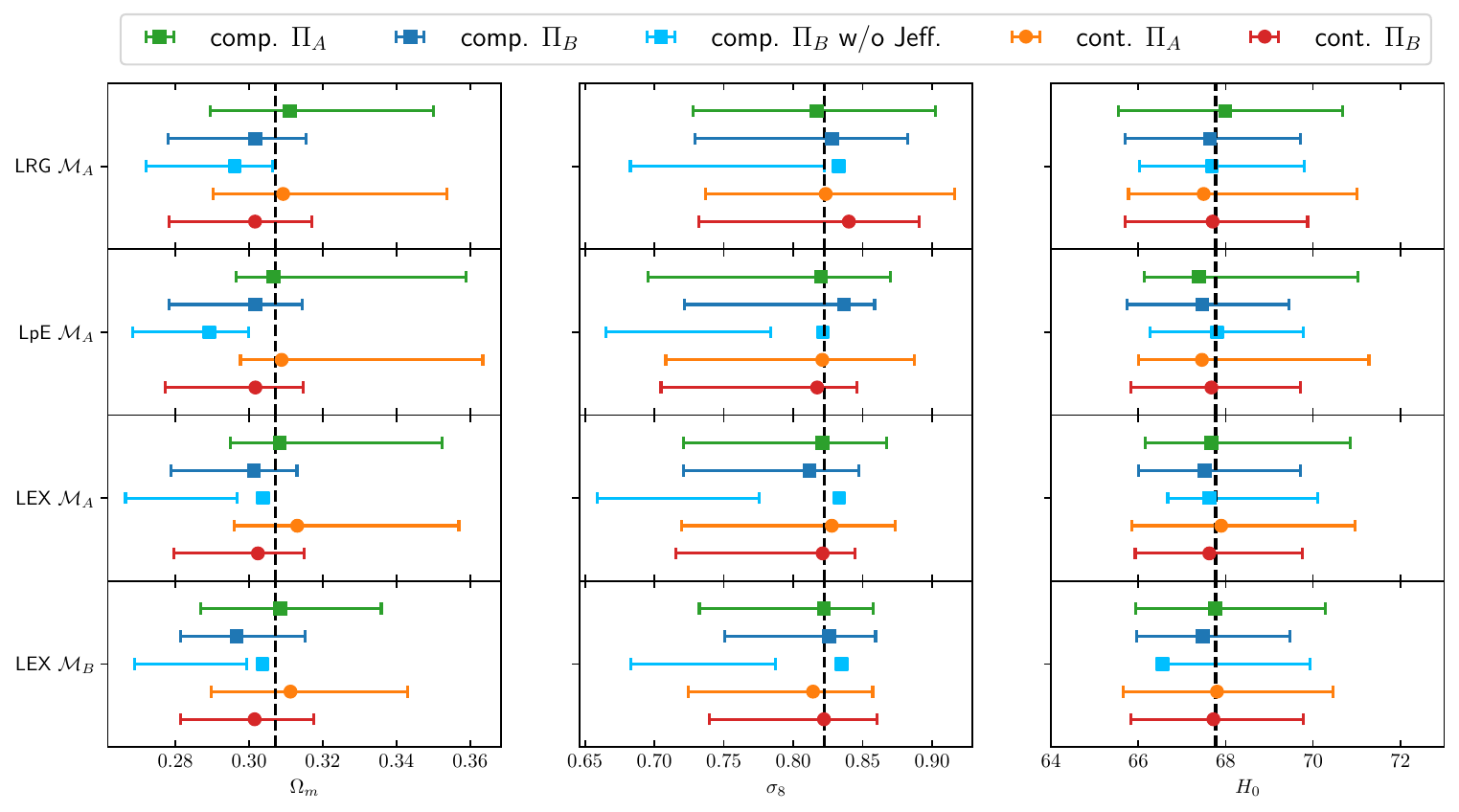}
    \caption{Summary of the best-fitting values and $68$ per cent credible intervals on cosmological parameters for different data sets and different prior choices derived from the EZmock analysis. The black dashed lines mark the fiducial cosmology of the EZmocks. $\mathcal{M}_A$ is the base model discussed in section~\ref{subsec:parameter-estimation}, while $\mathcal{M}_B$ fixes the stochastic terms in the cross power spectrum to zero. \textit{comp.} and \textit{cont.} represent the \textit{complete} EZmock and the \textit{contaminated} EZmock, respectively, and we use square and circle to mark MAPs corresponding to these two different EZmocks. We also show results employing a prior widely used in the literature, i.e. $\Pi_B$ prior without including the Jeffreys prior. In this case, especially when doing the multitracer analysis, although MAPs still remain unbiased, the fiducial cosmology is not covered by the $68$ per cent credible intervals.}
    \label{fig:EZmock-marg1D}
\end{figure*}
In this section we test our theory model and pipeline using 1000 realizations of multitracer EZmock catalogues \citep{Zhao:2020bib}. EZmock uses an approximate gravity solver based on the Zel'dovich approximation \citep{Zeldovich:19170} to generate the non-linear matter density field, and then sample galaxies based on an effective bias description. The fiducial cosmology used in this set of mocks is $\Omega_m=0.307115, \Omega_b=0.048206, h=0.6777, \sigma_8=0.8225, n_s=0.9611$. EZmock captures the survey geometry, redshift evolution and clustering properties of eBOSS DR16 LRG and ELG samples. The auto power spectrum measurements from EZmocks agree well\footnote{This is only true for ELG sample. There is a mismatch between mock and data at high $k$ for the CMASS+eBOSS sample because mocks for BOSS and eBOSS LRGs are calibrated separately and the cross correlation between those two samples is not precisely modelled.} with the actual data within $1\sigma$ for scales up to $k\sim 0.3\ h\,\mathrm{Mpc}^{-1}$. For the cross power spectrum, there are some discrepancies between mock and data for scales above $k=0.15\ h\,\mathrm{Mpc}^{-1}$, since LRGs and ELGs in the EZmock catalogues are populated independently (but share the same underlying matter density field), the small-scale cross correlation is not realistic in the EZmocks. As we will demonstrate later, the uncorrelated small scale noise leads to expected zero stochastic terms in our cross power spectrum theory model.

The EZmocks consist of three different sets of catalogues, the \textit{complete} mock, the \textit{contaminated} mock and the \textit{RIC} mock. The \textit{complete} mock and the \textit{RIC} mock are both free of observational systematics, but they use different random catalogues. The \textit{complete} mock uses a random catalogue constructed directly from the true selection function. While the \textit{RIC} mock employs the so-called \textit{shuffled} random catalogue, where the radial number density distribution $n(z)$ is inferred indirectly from the data, namely the mock galaxy catalogue, and thus it suffers from the RIC effect. The \textit{contaminated} mocks are injected with observational systematics, such as depth-dependent radial density, angular photometric systematics, fibre collision and redshift failure. They also suffer from the RIC effect. In this section, we focus only on the \textit{complete} mock, postponing the discussion of RIC effect and observational systematics to the next section.

\begin{figure}
    \includegraphics[width=\columnwidth]{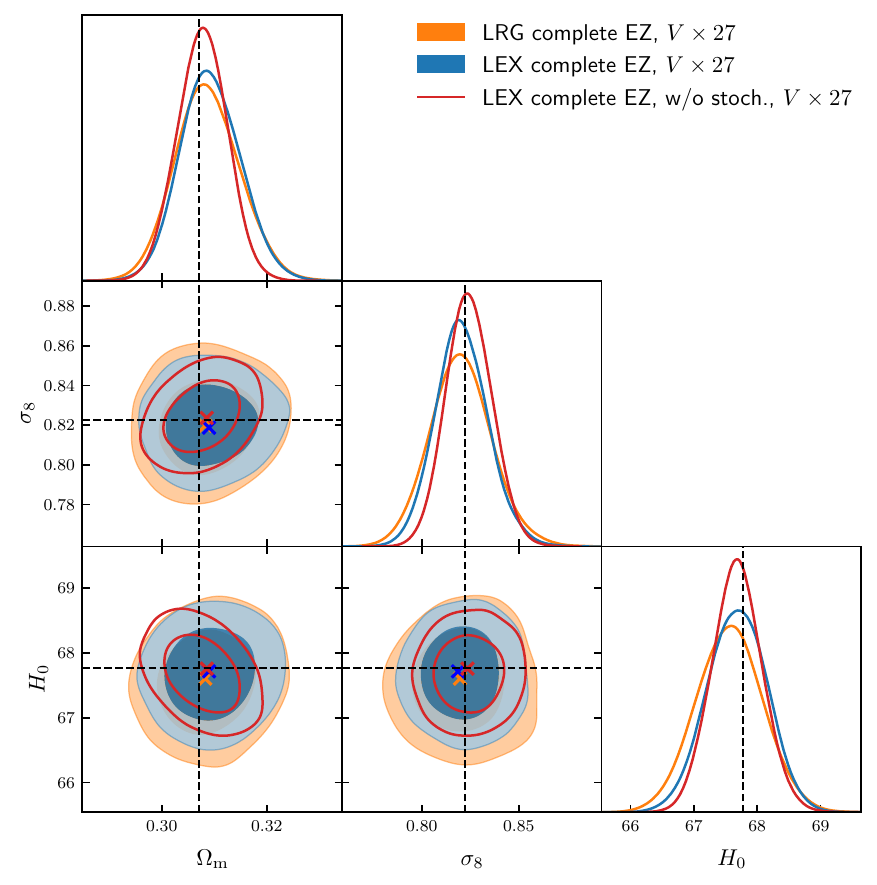}
    \caption{The 1D posterior alongside the $68$ per cent and $95$ per cent credible level contour plots for cosmological parameters derived from the \textit{complete} EZmock analysis using the uninformative $\Pi_A$ prior for the combined NGC and SGC samples. The covariance matrix are rescaled by 27. The black dashed lines mark the true cosmology and the cross marks denote the MAPs. The red contours represent the multitracer analysis making use of the LRG power spectrum, ELG power spectrum and the cross power spectrum between LRGs and ELGs but fixing the stochastic terms in cross to zero. For better visibility, the contours for ELG alone are not shown, which is also unbiased.}
    \label{fig:complete-EZmock-contour}
\end{figure}
\begin{figure*}
    \includegraphics[width=0.95\linewidth]{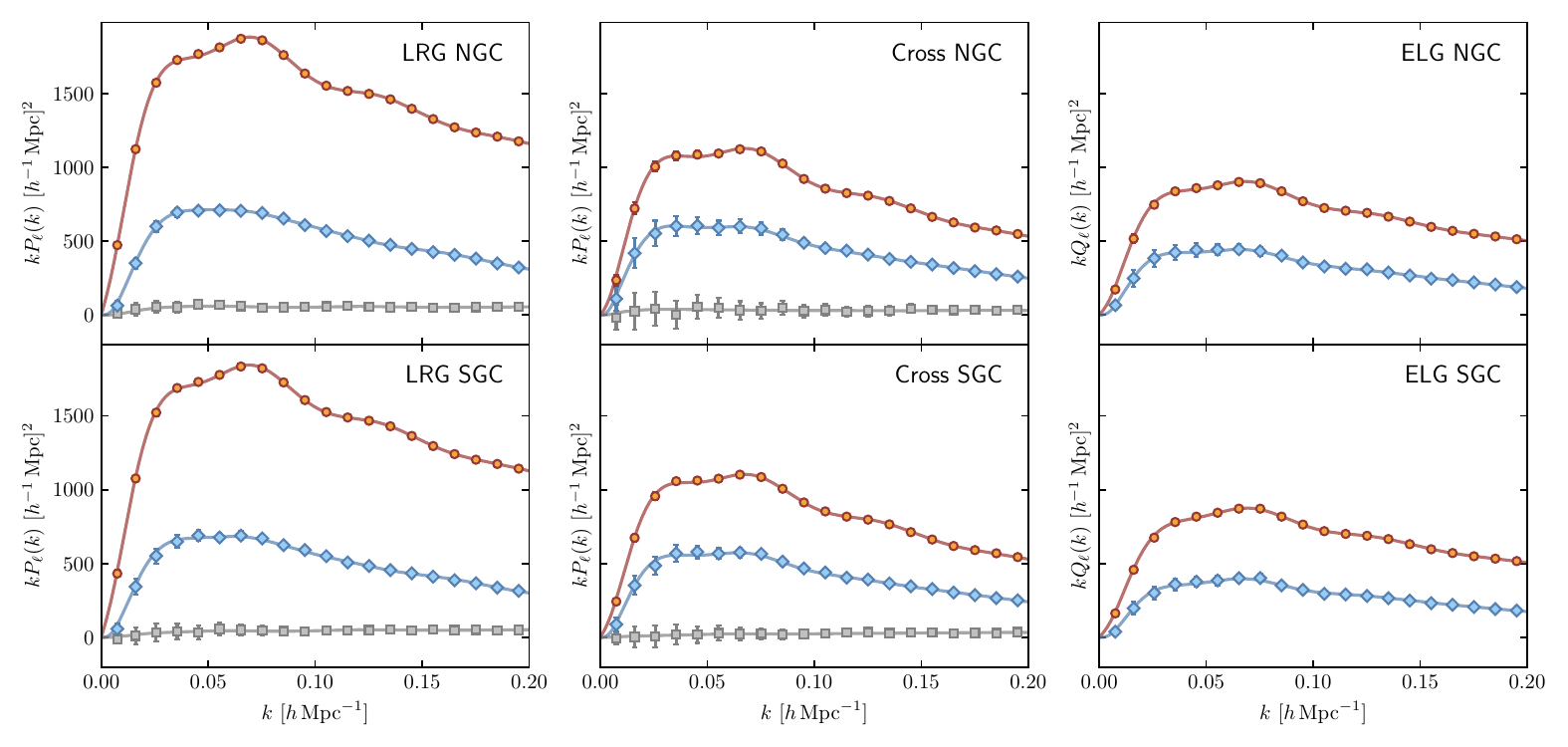}
    \caption{The best-fitting curves derived from the \textit{complete} EZmock multitracer analysis using the uninformative $\Pi_A$ prior. The error bars are taken from the diagonal part of the covariance matrix rescaled by a factor of 27. The red, blue and green lines correspond to $\ell=0$, $\ell=2$ and $\ell=4$ multipoles respectively.}
    \label{fig:complete-EZmock-bestfit}
\end{figure*}
\begin{table}
    \centering
    \caption{The mean values and the $68$ per cent credible intervals of stochastic terms in the cross power spectrum derived from the \textit{complete} EZmock analysis with covariance matrix rescaled by 27. The MAPs are given in brackets. We present results with cosmological parameters varied (the first two lines) and fixed (the last two lines). In all cases, the stochastic terms in the cross power spectrum are consistent with zero within $1\sigma$.}
    \begin{tabular}{ccc}
    \hline
    & $c_{\epsilon,0}^X$ & $c_{\epsilon, \mathrm{quad}}^X$ \\
    \hline\hline
    NGC w/ cosmo & $0.014^{+0.071}_{-0.11} (-0.003)$ & $-0.16\pm 0.18 (-0.108)$\\
    \hline
    SGC w/ cosmo & $0.002^{+0.094}_{-0.14} (-0.018)$ & $-0.11^{+0.23}_{-0.20} (-0.055)$ \\
    \hline
    NGC w/o cosmo & $0.013^{+0.060}_{-0.078} (0.031)$ & $-0.13^{+0.18}_{-0.16} (-0.129)$ \\
    \hline
    SGC w/o cosmo & $0.003^{+0.081}_{-0.11} (0.014)$ & $-0.08^{+0.22}_{-0.19} (-0.063)$ \\
    \hline
    \end{tabular}
    \label{tab:EZmock-stochastic-terms}
\end{table}
We fit to the mean power spectrum multipoles averaged over 1000 realizations of the \textit{complete} EZmocks. To suppress the volume effect and highlight any bias induced by modelling error, we rescale the covariance matrix by 27\footnote{The choice of rescaling factor is arbitrary and we take this number from \citet{Ivanov:2021zmi}.}. The contour plot and the best-fitting curves are shown in Fig.~\ref{fig:complete-EZmock-contour} and Fig.~\ref{fig:complete-EZmock-bestfit}, respectively. Both the single-tracer and the multitracer analysis yield unbiased results, and their MAPs are consistent. This proves our treatment of survey window convolution is correct in the pipeline and the assumption of ignoring the redshift evolution of EFT parameters in the cross power spectrum does not introduce noticeable bias. In addition, the cross power spectrum stochastic terms are measured to be zero within $1\sigma$ as shown in Table~\ref{tab:EZmock-stochastic-terms}, which is consistent with the independent galaxy population procedure when constructing EZmock catalogues.

Using rescaled covariance matrices we have verified our pipeline is unbiased. Returning to the case with normal covariance, we investigate how the remaining volume effect might alter our results. The marginalized cosmological results are summarized in Fig.~\ref{fig:EZmock-marg1D}. For brevity, the base model discussed in section~\ref{subsec:parameter-estimation} is referred to as $\mathcal{M}_A$, and the model with stochastic terms fixed to zero in the cross power spectrum is named as $\mathcal{M}_B$. Similar to what we have discussed in the last section, both $\Pi_A$ and $\Pi_B$ recover the fiducial cosmology within $1\sigma$. The difference between those two priors manifests mainly in parameter $\Omega_m$ for which the more conservative prior $\Pi_A$ favours a higher value of $\Omega_m$ while the more aggressive prior $\Pi_B$ favours a slightly smaller $\Omega_m$\footnote{The argument relies on the specific data set and hyperparameters we choose in Table~\ref{tab:nuisance-prior}.}. A more interesting case is $\Pi_B$ without the Jeffreys prior, i.e. the classic Gaussian prior widely used in the literature. The best-fitting values are determined by minimizing the full posterior\footnote{In practice, when sampling the marginalized posterior, we save the full posterior $\mathcal{P}_\mathrm{full}$ at each step and select the point with largest $\mathcal{P}_\mathrm{full}$ as the best-fit to save computational time.} searching through the non-linear parameter subspace with linear parameters computed by equation~(\ref{eq:bG_best}) following \citet{DAmico:2020kxu} and \citet{Simon:2022csv}.  In this case, the best-fitting values may not fall within the $68$ per cent credible interval particularly when combining with ELG data. This is because the ELG data alone is weak (using the chained power spectrum also loses information) and suffers from a very strong volume effect. Therefore, using the Jeffreys prior is crucial for the eBOSS multitracer analysis. On the other hand, the best-fitting values are consistent with the truth.

Compared to the single-tracer analysis, the multitracer analysis improves cosmological constraints by $\sim10\%$ (or more if stochastic terms in the cross power spectrum are further fixed to zero). However, the improvement is underestimated, because LRG samples cover a much larger sky area than ELG samples and thus the cosmological information is dominated by the LRGs. If the multitracer analysis is performed with a comparable footprint for both samples, the improvement would be at $\sim30\%$ level. We refer the readers to Appendix~\ref{appendix:a} for detailed discussion.

\subsection{Contaminated EZmock}
\label{subsec:contaminated-mock}
\begin{figure}
    \includegraphics[width=\columnwidth]{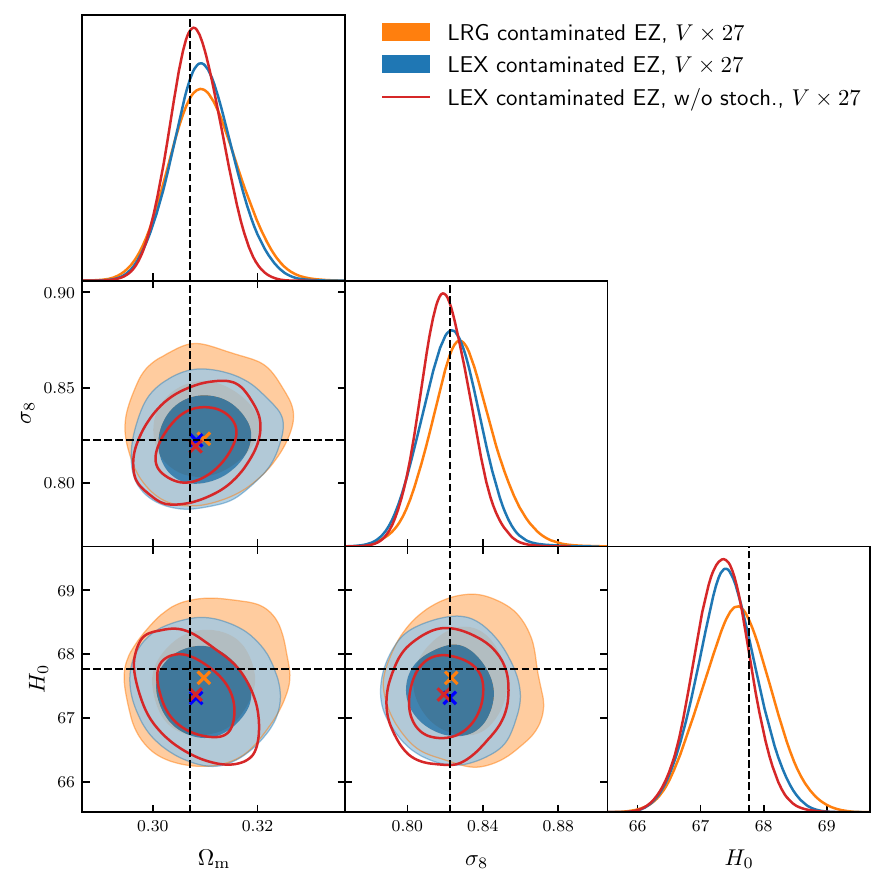}
    \caption{Similar to Fig.~\ref{fig:complete-EZmock-contour}, but derived from the \textit{contaminated} EZmocks.}
    \label{fig:contaminated-EZmock-contour}
\end{figure}
\begin{figure*}
    \includegraphics[width=0.95\linewidth]{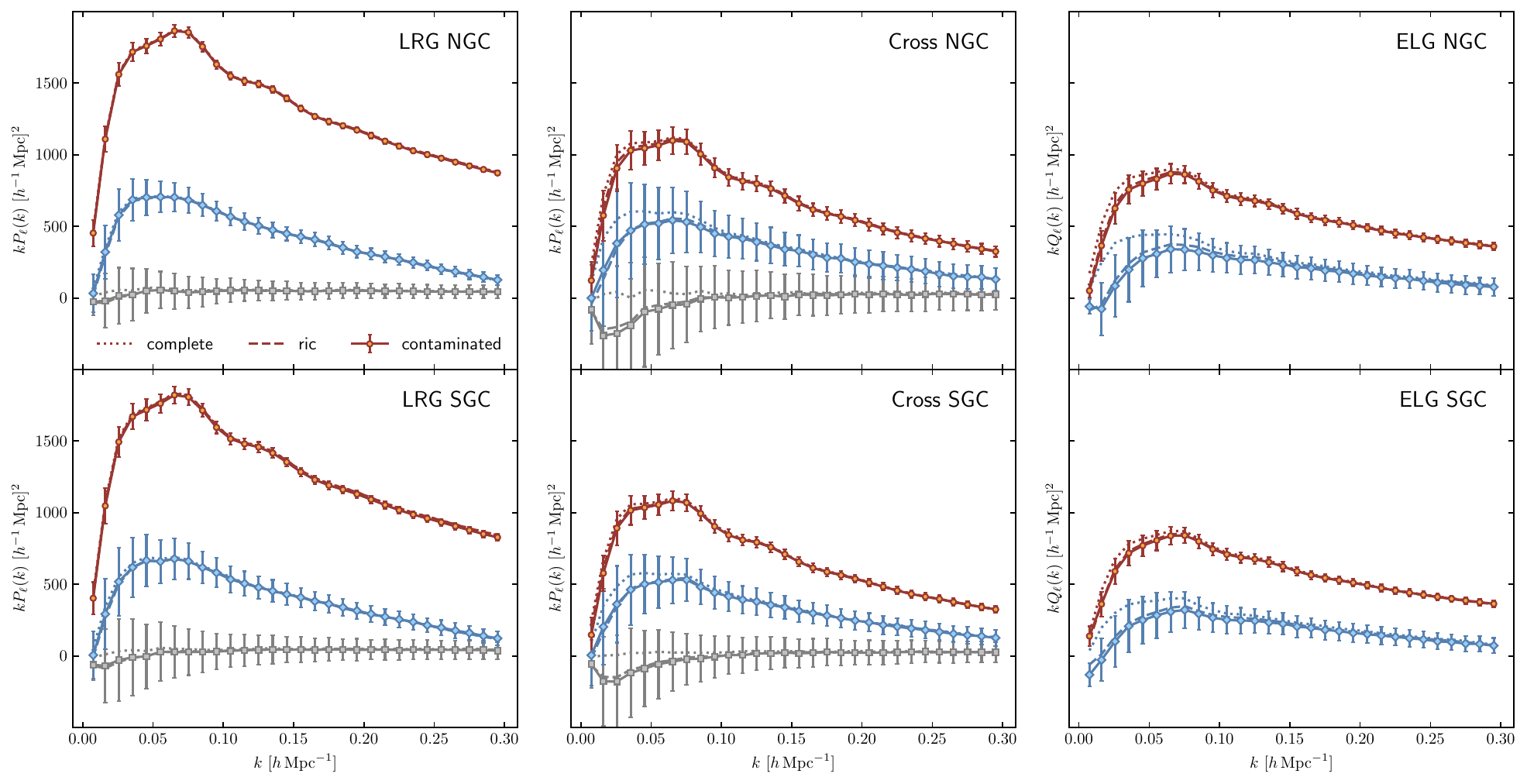}
    \caption{The power spectrum multipoles measured from EZmocks averaged over 1000 realizations. The red, blue and green lines correspond to $\ell=0$, $\ell=2$ and $\ell=4$ multipoles respectively.}
    \label{fig:EZmock-pk}
\end{figure*}
In order to study the impact of systematics, we repeat the analysis in the last section, substituting the \textit{complete} EZmocks with the \textit{contaminated} EZmocks.

Fitting results with covariance matrices rescaled by 27 are shown in Fig.~\ref{fig:contaminated-EZmock-contour}. After including ELG and the cross power spectrum, we see a minor shift (less than $1\sigma$) in $H_0$, indicating our systematics corrections are not perfect. This is expected, as the chained power spectrum discussed in section~\ref{subsec:chained-pk} does not fully eliminate the angular systematics in the ELG power spectrum as seen by comparing the \textit{contaminated} curves to the \textit{RIC} curves measured from EZmocks (see Fig.~\ref{fig:EZmock-pk}). There are also systematics in the cross power spectrum at large scales although a common belief is that independent systematics in different samples are not correlated and do not appear in the cross power spectrum. However, given their small magnitude, they do not affect our final results.

The marginalized cosmological constraints with normal unrescaled covariance matrices are displayed in Fig.~\ref{fig:EZmock-marg1D}. The best-fitting values and credible intervals are consistent with results derived from the \textit{complete} EZmocks, and thus validate our pipeline for a realistic survey.

\section{Results}
\label{sec:results}
\begin{figure*}
    \includegraphics[width=0.95\linewidth]{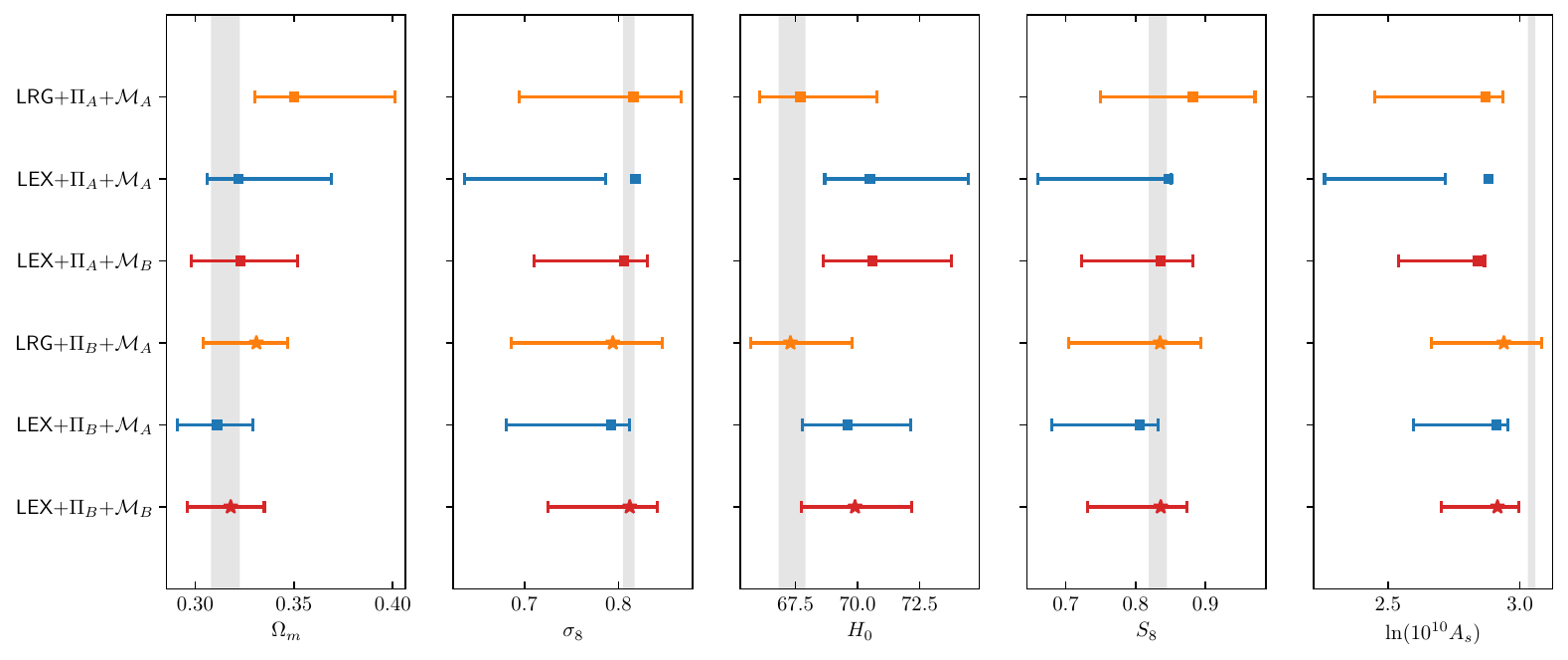}
    \caption{The best-fitting values and the $68$ per cent credible intervals of cosmological parameters measured from eBOSS DR16 data for different data set combinations and different priors and configurations. The black bands show the $1\sigma$ Planck~2018 TT,TE,EE+lowE+lensing baseline results. The best-fitting values from our baseline analysis are marked with a star symbol.}
    \label{fig:data-compare-Planck18}
\end{figure*}
\begin{table*}
    \centering
    \caption{Cosmological parameters measured from eBOSS DR16 data for different priors, configurations and data set combinations. The underlined cases represent our primary results. There are two or three rows for each case, from top to bottom: the best-fitting value, the $68$ per cent credible interval and the multitracer improvement relative to LRG alone with the same prior (positive improvements are highlighted in bold font). We also show Planck~2018 TT,TE,EE+lowE+lensing baseline results for comparison. The corresponding $\chi^2$ value is calculated at the best-fitting point.}
    \begin{tabular}{ccccccc}
    \hline
    case & $\Omega_m$ & $\sigma_8$ & $H_0$ & $S_8$ & $\ln(10^{10}A_s)$ & $\chi^2/\mathrm{d.o.f.}$ \\
    \hline\hline
    \multirow{2}{8em}{\centering LRG+$\Pi_A$+$\mathcal{M}_A$} & $0.350$ & $0.816$ & $67.7$ & $0.882$ & $2.896$ & \multirow{2}{7em}{\centering $106.1/(108-20)$} \\
    & $0.367^{+0.034}_{-0.037}$ & $0.788^{+0.079}_{-0.094}$ & $68.6^{+2.2}_{-2.5}$ & $0.871^{+0.099}_{-0.12}$ & $2.72\pm 0.24$ & \\
    \hline
    \multirow{3}{8em}{\centering LEX+$\Pi_A$+$\mathcal{M}_A$}  & $0.322$  & $0.818$ & $70.5$ & $0.847$ & $2.881$ & \multirow{3}{7em}{\centering $288.1/(284-40)$}\\
    & $0.341^{+0.028}_{-0.035}$  & $0.717^{+0.069}_{-0.081}$ & $71.6\pm 3.1$ & $0.764^{+0.086}_{-0.10}$ & $2.49\pm 0.23$ &\\
    & $\mathbf{+10.1\%}$ & $\mathbf{+13.5\%}$ & $-22.0\%$ & $\mathbf{+13.7\%}$ & $\mathbf{+5.5\%}$ &\\
    \hline
    \multirow{3}{8em}{\centering LEX+$\Pi_A$+$\mathcal{M}_B$}  & $0.323$ & $0.806$ & $70.6$ & $0.836$ & $2.842$ & \multirow{3}{7em}{\centering $293.0/(284-36)$}\\
    & $0.327^{+0.025}_{-0.029}$ & $0.772\pm 0.061$ & $71.5^{+2.3}_{-2.8}$ & $0.805\pm 0.080$ & $2.70\pm 0.16$ &\\
    & $\mathbf{+24.9\%}$ & $\mathbf{+30.0\%}$ & $-8.9\%$ & $\mathbf{+28.0\%}$ & $\mathbf{+32.7\%}$ &\\
    \hline\hline
    \multirow{2}{8em}{\centering \underline{LRG+$\Pi_B$+$\mathcal{M}_A$}}  & $0.331$ & $0.794$ & $67.3$ & $0.835$ & $2.939$ & \multirow{2}{7em}{\centering $108.8/(108-20)$}\\
    & $0.329^{+0.019}_{-0.024}$ & $0.772^{+0.075}_{-0.085}$ & $67.7\pm 2.1$ & $0.808^{+0.084}_{-0.10}$ & $2.87\pm 0.21$ & \\
    \hline
    \multirow{3}{8em}{\centering LEX+$\Pi_B$+$\mathcal{M}_A$}  & $0.311$ & $0.792$ & $69.6$ & $0.806$ & $2.911$ & \multirow{3}{7em}{\centering $295.3/(284-40)$}\\
    & $0.312^{+0.017}_{-0.021}$  & $0.747\pm 0.067$ & $70.1\pm 2.2$ & $0.762^{+0.070}_{-0.081}$ & $2.76^{+0.19}_{-0.17}$ &\\
    & $\mathbf{+12.2\%}$ & $\mathbf{+18.1\%}$ & $-7.4\%$ & $\mathbf{+19.8\%}$ & $\mathbf{+14.8\%}$ &\\
    \hline
    \multirow{3}{8em}{\centering \underline{LEX+$\Pi_B$+$\mathcal{M}_B$}} & $0.318$ & $0.812$ & $69.9$ & $0.836$ & $2.915$ & \multirow{3}{7em}{\centering $297.7/(284-36)$}\\
    & $0.317^{+0.017}_{-0.021}$ & $0.787^{+0.055}_{-0.062}$ & $70.0\pm 2.3$ & $0.809^{+0.064}_{-0.078}$ & $2.85\pm 0.15$ &\\
    & $\mathbf{+10.8\%}$ & $\mathbf{+27.3\%}$ & $-9.7\%$ & $\mathbf{+24.6\%}$ & $\mathbf{+29.3\%}$ &\\
    \hline\hline
    \multirow{2}{8em}{\centering Planck~2018} & $0.3158$ & $0.8120$ & $67.32$ & $0.8331$ & $3.0448$ & \multirow{2}{7em}{\centering -}\\
     & $0.3153\pm0.0073$ & $0.8111\pm0.0060$ & $67.36\pm0.54$ & $0.832\pm0.013$ & $3.044\pm0.014$ &\\
    \hline
    \end{tabular}
    \label{tab:eBOSS-data-cosmo-summary}
\end{table*}
Our final cosmological parameter constraints are summarized in Table~\ref{tab:eBOSS-data-cosmo-summary} and Fig.~\ref{fig:data-compare-Planck18}, making use of the full eBOSS DR16 LRG and ELG samples in both NGC and SGC. The best-fitting curves for the LEX data set, employing the $\Pi_B$ prior and assuming zero stochastic terms in the cross power spectrum, are displayed in Fig.~\ref{fig:data-fit}. We refer the readers to Appendix~\ref{appendix:b} for the full parameter constraints.

Our multitracer and single-tracer analysis results are in general consistent. However, including ELGs and the cross power spectrum shifts $H_0$ to a higher value by $\sim1\sigma$. As we have discussed in section~\ref{subsec:contaminated-mock}, the impact of remaining angular systematics is small, and considering the weak constraints provided by the ELG auto power spectrum, we suspect the shift is caused by the remaining RIC in the cross power spectrum. However, the amount of shift is only at $1\sigma$ level. The multitracer analysis improves the constraints on $\Omega_m$ by $\sim10\%- 25\%$, $\sigma_8$ by $\sim13\%- 30\%$ depending on the choice of priors and configurations while the constraints on $H_0$ degrades, which is due to the slight degeneracy direction rotation in the $\Omega_m\text{--}H_0$ plane.

These results are also consistent across different configurations. However, we find that in the case of multitracer analysis with the uninformative prior $\Pi_A$ and model $\mathcal{M}_A$ which includes the stochastic terms in the cross power spectrum, the MAP does not fall within the $68$ per cent credible interval, indicating a severe remaining volume effect. Although we have demonstrated that Jeffreys prior can mitigate the volume effect in section~\ref{subsec:complete-mock}, when applying this method to realistic data, removing the volume effect related to linear parameters is not sufficient, making the reported $68$ per cent credible interval strongly biased, especially for $\ln(10^{10}A_s)$. The discrepancy may stem from the different degeneracy shape between $c_2$ and $\sigma_8$, which appears rectangular when fitting to EZmocks but triangular with the realistic eBOSS data (see Fig.~\ref{fig:data-contour-full}), making our volume effect mitigation approach less efficient. Additionally, compared to the single LRG analysis, the constraints on $c_2^\mathrm{LRG}$ derived from multitracer analysis is weaker, leading to a stronger remaining volume effect. Therefore, we select $\Pi_B$ (Jeffreys prior + Gaussian prior) as our baseline prior, which effectively suppresses the remaining volume effect. We note that there is a slight shift in $\Omega_m$ after applying the Gaussian prior ($\Pi_B$), but as we demonstrated in section~\ref{subsec:complete-mock}, this shift is due to the exclusion of unphysical regions in the parameter space and using $\Pi_B$ can still recover the truth within $1\sigma$.

As discussed in previous sections, we included the stochastic terms in the cross power spectrum ($\mathcal{M}_A$) to account for the possible noise correlation between LRG and ELG samples. However, after comparing the results of $\mathcal{M}_A$ and $\mathcal{M}_B$, we find that the best-fitting cosmological parameters inferred from these two models agree well within $1\sigma$. Also the reduced chi-square $\chi^2/\mathrm{d.o.f.}$ is close. We conclude that the precision of eBOSS data cannot distinguish between these two models. Thus, we will report LRG+$\Pi_B$+$\mathcal{M}_A$ and LEX+$\Pi_B$+$\mathcal{M}_B$ as the main results in this work. The constraints on $\Omega_m$ and $\sigma_8$ are improved by $10.8\%$ and $27.3\%$ respectively, while the constraint on $H_0$ degrades by $9.7\%$. The Figure of Merit (FoM) of cosmological parameters $\mathrm{FoM}=1/\sqrt{\det\mathrm{Cov}(\Omega_m, \sigma_8, H_0, \omega_b)}$ is improved by $62.7\%$ compared to the single LRG analysis. The improvement on $\sigma_8$ is close to the multitracer analysis with comparable footprints presented in Appendix~\ref{appendix:a}, demonstrating the efficiency of this method even with the limited overlapping area in realistic eBOSS data.

In order to compare our results with other analysis like Planck~2018, we use $(\mu_1 - \mu_2) / \sqrt{\sigma_1^2 + \sigma_2^2}$ as the $\sigma$-deviation metric following \citet{Simon:2022csv}, where $\mu_i$ denotes the mean and $\sigma_i$ represents the error bar read from the $68$ per cent credible interval. We find all cosmological parameters are consistent with Planck at $\lesssim 1.3\sigma$ level. In particular, in our analysis
\begin{align}
    &H_0 = 67.7\pm2.1,\quad S_8=0.808^{+0.084}_{-0.10}\quad\text{(LRG),}\\
    &H_0 = 70.0\pm2.3,\quad S_8=0.809^{+0.064}_{-0.078}\quad\text{(ALL),}
\end{align}
and we find no $H_0$ and $S_8$ tension between the eBOSS large-scale structure analysis and Planck~2018. The value of $H_0$ inferred from our analysis is $2.3\sigma$ lower than that reported by SH0ES \citep{Riess:2021jrx} for the LRG analysis and $1.2\sigma$ lower for the LEX analysis. The inferred value of $S_8$ is in general higher compared to the weak lensing measurement \citep{KiDS:2020suj,DES:2021wwk,HSC1:2023,HSC2:2023,HSC3:2023,LSSTDarkEnergyScience:2022amt,DES-KIDS:2023}, but statistically consistent at $\lesssim 0.6\sigma$ level.

In order to compare our results with previous template fitting analysis, we convert cosmological parameters to the standard AP parameters and RSD parameter. The official eBOSS collaboration paper \citep{Gil-Marin:2020bct} reported $D_H(z_\mathrm{eff}=0.698)/r_\mathrm{drag}=19.77\pm 0.47, D_M(z_\mathrm{eff}=0.698)/r_\mathrm{drag}=17.65\pm0.30, f\sigma_8(z_\mathrm{eff}=0.698)=0.473\pm0.044$ for LRG samples combining Fourier space and configuration space BAO and full-shape information. Our analysis yields
\begin{align}
    &D_H(z_\mathrm{eff}=0.698)/r_\mathrm{drag}=20.24\pm0.28\\
    &D_M(z_\mathrm{eff}=0.698)/r_\mathrm{drag}=17.61\pm0.35\\
    &f\sigma_8(z_\mathrm{eff}=0.698)=0.442^{+0.043}_{-0.050}\text{,}
\end{align}
which are in agreement with the official eBOSS results at $\lesssim 0.9\sigma$ level. \citet{Wang:2020tje} performed a multitracer analysis at $z_\mathrm{eff}=0.77$ in configuration space based on template fitting, reporting $D_H(z_\mathrm{eff}=0.77)/r_\mathrm{drag}=19.64\pm 0.57, D_M(z_\mathrm{eff}=0.77)/r_\mathrm{drag}=18.85\pm0.38, f\sigma_8(z_\mathrm{eff}=0.77)=0.432\pm0.038$ and $f\sigma_8(z_\mathrm{eff}=0.77)=0.472\pm0.043$ for LRG alone. Our analysis yields
\begin{align}
    &D_H(z_\mathrm{eff}=0.77)/r_\mathrm{drag}=19.08\pm0.27\\
    &D_M(z_\mathrm{eff}=0.77)/r_\mathrm{drag}=18.67\pm0.39\\
    &f\sigma_8(z_\mathrm{eff}=0.77)=0.444^{+0.033}_{-0.037}\\
    &f\sigma_8(z_\mathrm{eff}=0.77)=0.436^{+0.042}_{-0.049}\quad(\mathrm{LRG})\text{,}
\end{align}
which are statistically consistent at $\lesssim 0.9\sigma$ level. By performing the multitracer analysis, the statistical uncertainty of $f\sigma_8$ derived from our EFT analysis is reduced by $27.2\%$, showing a better improvement than $11.6\%$ based on template fitting.

\section{Conclusion and discussions}
\label{sec:conclusion}
In this work, we have performed a multitracer full-shape analysis of eBOSS DR16 LRG and ELG samples based on the effective field theory of large-scale structure (EFTofLSS).

We examined the prior effect related to EFT parameters, utilising one noiseless synthetic mock created from the theory code and 1000 eBOSS EZmocks. We detected a strong volume effect when analysing samples especially with a uniform prior, which shifts the best fit value of $\sigma_8$ to a lower value. This effect is most significant when combined with ELG data due to its weak constraining power. We demonstrated that the use of a Jeffreys prior can mitigate the volume effect, exactly cancelling out the effect related to linear EFT parameters. Additionally, we explore a combination of Jeffreys prior and Gaussian prior, which achieves similar performance but avoids including unphysical regions in the parameter space by limiting the magnitude of the linear nuisance parameters. However, we note that the use of Gaussian prior requires choosing related hyperparameters and provides non-negligible additional information even with a DESI-like volume. We expect that calibrations using simulations can set reasonable priors for these parameters.

Before applying to the analysis of real data, we verified that our multitracer pipeline can yield unbiased cosmological constraints using both the \textit{complete} and \textit{contaminated} EZmock catalogues. We then find that when combined with the BBN prior and the spectral tilt $n_s$ fixed to Planck value, our multitracer analysis of eBOSS DR16 data measures $H_0=70.0\pm2.3$, $\Omega_m=0.317^{+0.017}_{-0.021}$, $\sigma_8=0.787_{-0.062}^{+0.055}$ and $S_8=0.809_{-0.078}^{+0.064}$, consistent with the Planck~2018 results. We find the figure of merit of cosmological parameters is improved by $62.7\%$ compared to the single LRG analysis. Specifically, the statistical uncertainty of $\sigma_8$ is reduced by $27.3\%$ in our baseline analysis.

Furthermore, we performed a DESI-like multitracer forecast for samples with comparable sky coverages based on EZmocks. We find multitracer analysis can improve the constraints on $\Omega_m$ and $H_0$ by $\sim 25\%$, and $\sigma_8$ by $\sim 15\%$ at $k_\mathrm{max}=0.20\ h\,\mathrm{Mpc}^{-1}$ regardless of the prior we use. If we assume vanishing stochastic terms in the cross power spectrum, the improvements are further boosted to $\sim 30\%$ and $\sim 25\%$. This improvement is not directly related to the cosmic variance cancellation but comes from breaking degeneracies between stochastic terms and cosmological parameters, which merits future investigation with the aid of $N$-body based mocks and multitracer HOD models.

The tightest cosmological constraints so far in the large-scale structure perturbative full-shape analyses were obtained from a joint analysis of power spectrum and bispectrum \citep{Philcox:2022frc,Philcox:2021kcw,DAmico:2022osl,Ivanov:2023qzb}. Our results are in agreement with these analyses at $\lesssim 1\sigma$ level. Compared to including bispectrum, the multitracer analysis achieves comparable improvement on $\sigma_8$. It is worth noting that the $\sigma_8$ derived from our analysis is $1\sigma$ higher than analyses using the ``East-coast'' prior, which is possibly caused by the volume effect \citep{Simon:2022lde}. Also a recent paper \citep{donald-mccann:2023} points out that the application of Jeffreys prior, as employed in this work, can mitigate the volume effect and leads to a higher $\ln(10^{10}A_s)$ when analysing BOSS data, improving the agreement with Planck~2018 results.

The methods developed in this work are applicable to the forthcoming DESI survey \citep{DESI:2016fyo}, which observes multiple tracers at higher redshifts with higher number densities. It would be interesting to explore the multitracer improvement beyond the standard $\Lambda$CDM model, such as massive neutrinos \citep{Ivanov:2019hqk,Aviles:2021que} and primordial non-Gaussianity \citep{Dalal:2007cu,Slosar:2008hx,Sullivan:2023qjr}. It would also be interesting to extend our analysis to higher order statistics, such as galaxy bispectrum \citep{Yamauchi:2016wuc,Karagiannis:2023lsj}, or non-standard statistics, e.g. post-reconstructed power spectrum \citep{Hikage:2017tmm,Hikage:2019ihj,Hikage:2020fte,Wang:2022nlx}, marked power spectrum \citep{Philcox:2020fqx}, galaxy voids \citep{Nadathur:2019mct,Zhao:2021ahg}, etc.

\section*{Acknowledgements}
We thank Pierre Zhang for his help with \code{PyBird} code, Cheng Zhao for providing EZmock catalogues and useful discussion on the angular incompleteness and normalization issue, Harry Desmond and Boryana Hadzhiyska for their valuable comments on the Jeffreys prior, Nathan Findlay for reading the draft and valuable comments. We are also grateful to Mikhail M. Ivanov, Giovanni Cabass, Arnaud de Mattia and Haruki Ebina for useful discussion. RZ, XM, YF, WZ, YW and GBZ are supported by NSFC grants 11925303, 11890691, and by the CAS Project for Young Scientists in Basic Research (No. YSBR-092). RZ is also supported by the Chinese Scholarship Council (CSC) and the University of Portsmouth. YW is also supported by NSFC Grants 12273048, by National Key R\&D Program of China No. 2022YFF0503404, by the Youth Innovation Promotion Association CAS, and by the Nebula Talents Program of NAOC. GBZ is also supported by science research grants from the China Manned Space Project with No. CMS-CSST-2021-B01, and the New Cornerstone Science Foundation through the XPLORER prize. JD-M and RG are supported by STFC studentships. KK is supported by the STFC grant ST/W001225/1. The authors acknowledge the Beijing Super Cloud Center (\href{https://www.blsc.cn/}{BSCC}) for providing HPC resources that have contributed to the research results reported within this paper. For the purpose of open access, the authors have applied a Creative Commons Attribution (CC BY) licence to any Author Accepted Manuscript version arising.
\section*{Data Availability}
All data measurements and pipeline code will be public at \url{https://github.com/zhaoruiyang98/eftpipe} upon publication of this work. Any additional products are available upon reasonable request to the corresponding author.




\bibliographystyle{mnras}
\bibliography{main} 




\appendix

\section{multitracer analysis with comparable footprints}
\label{appendix:a}
\begin{figure}
    \includegraphics[width=\columnwidth]{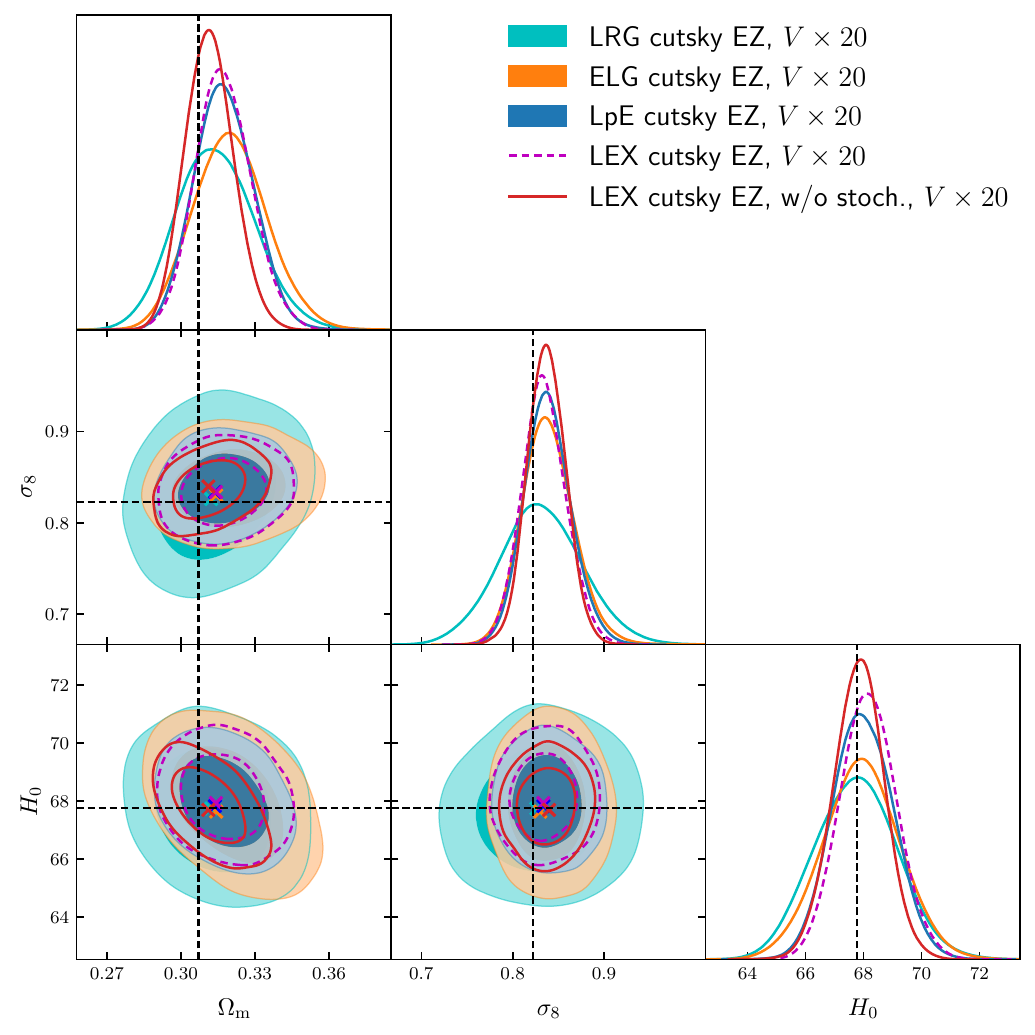}
    \caption{Similar to Fig.~\ref{fig:complete-EZmock-contour}, but for cutsky mocks. The blue filled contours show the constraints derived from two auto power spectrum.}
    \label{fig:cutsky-EZmock-contour}
\end{figure}
\begin{figure}
    \includegraphics[width=\columnwidth]{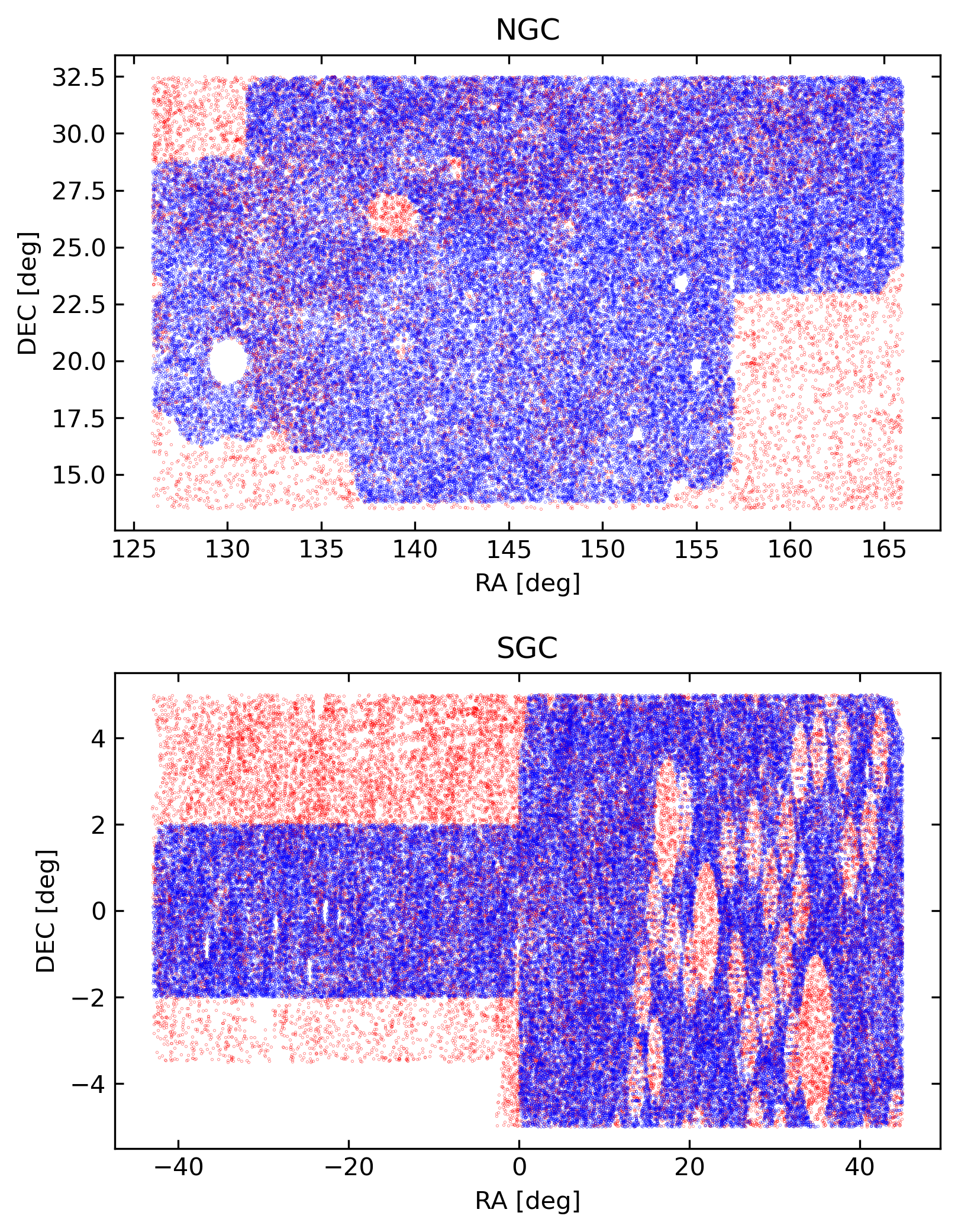}
    \caption{The cutsky footprint of LRG (red) and ELG (blue) samples in the North Galactic Cap (NGC) and South Galactic Cap (SGC).}
    \label{fig:footprint-cutsky}
\end{figure}
\begin{figure}
    \includegraphics[width=\columnwidth]{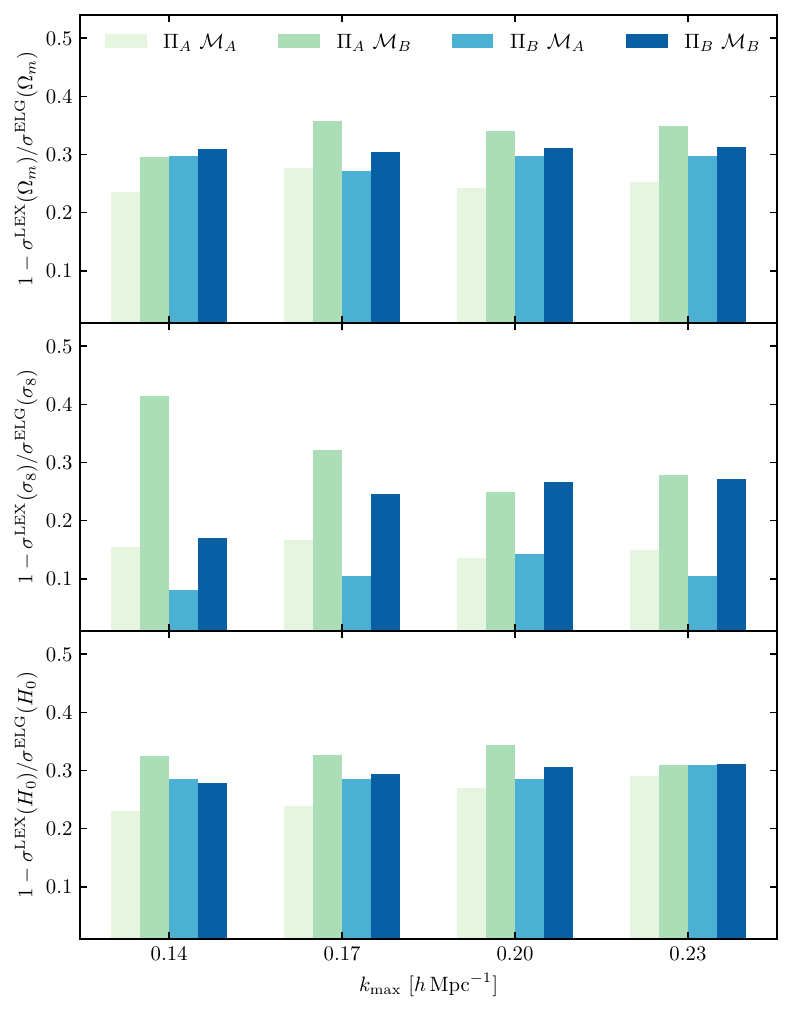}
    \caption{The bar plot shows the multitracer analysis (LEX) improvements relative to the single tracer analysis (ELG) on cosmological parameters for different prior choices ($\Pi$) and configurations ($\mathcal{M}$) at different $k_\mathrm{max}$. We use the same notations as used in Fig.~\ref{fig:EZmock-marg1D}.}
    \label{fig:cutsky-EZmock-bar}
\end{figure}
\begin{figure*}
    \includegraphics[width=0.95\linewidth]{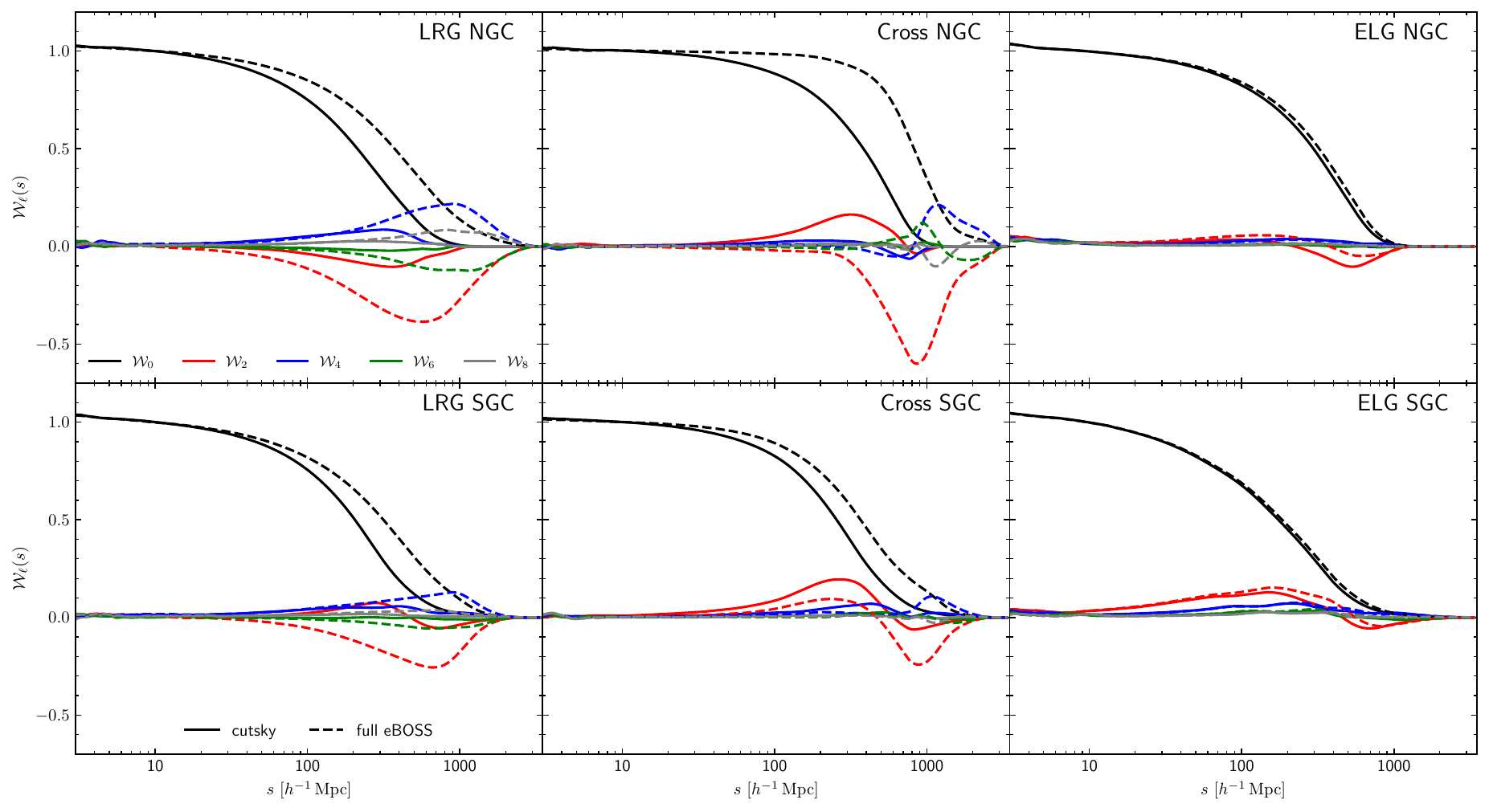}
    \caption{The window function multipoles similar to Fig.~\ref{fig:window}, but measured from the \textit{complete} EZmock random catalogue. The solid lines show the cutsky window functions, and the dashed lines correspond to the full eBOSS footprint for comparison.}
    \label{fig:window-cutsky}
\end{figure*}
\begin{figure*}
    \includegraphics[width=0.95\linewidth]{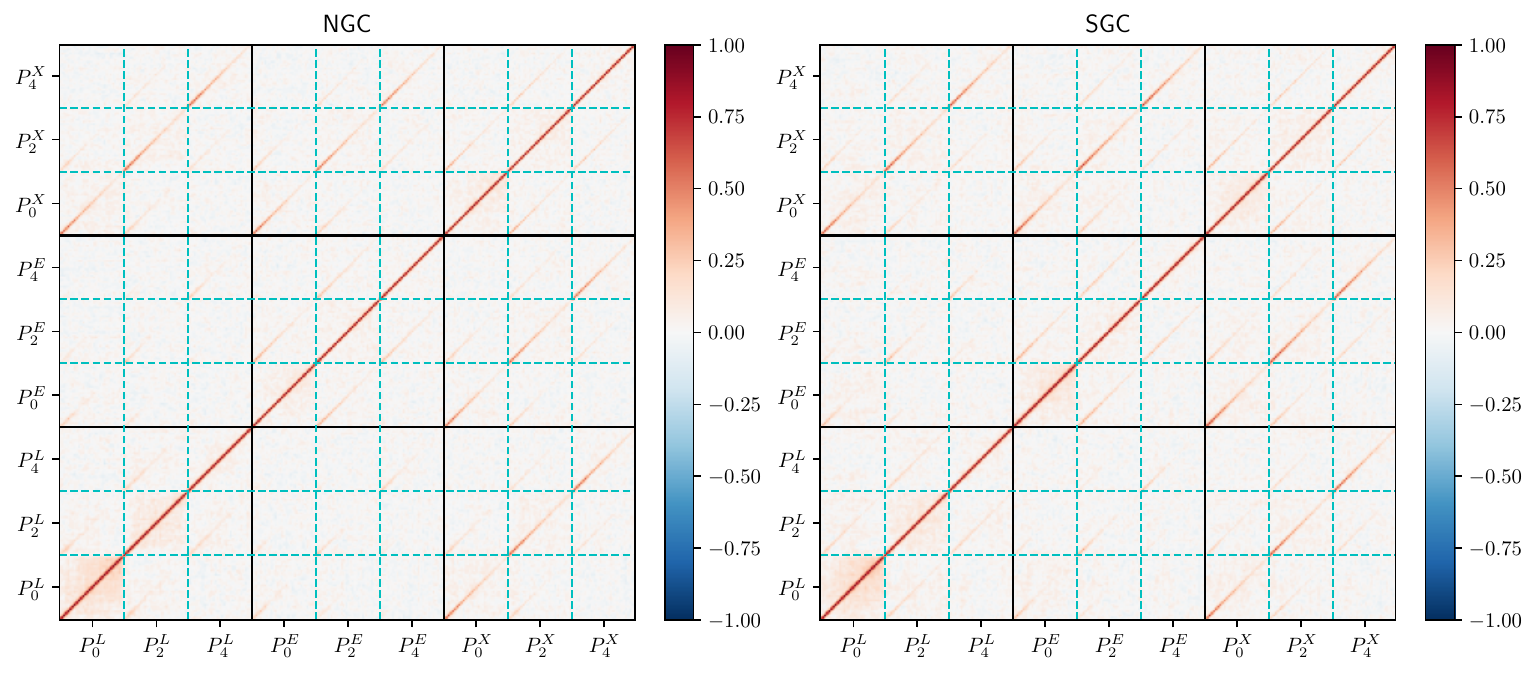}
    \caption{The correlation matrix similar to Fig.~\ref{fig:corr_full}, but measured from the cutsky \textit{complete} EZmocks, and here we show the power spectrum multipoles for LRG, ELG and the cross. The non-diagonal correlations are enhanced because cutsky samples are more overlapped.}
    \label{fig:corr_cutsky}
\end{figure*}
In order to provide a fair comparison between the single-tracer and multitracer analysis, we cut the footprint of LRG samples so that both LRGs and ELGs have comparable sky coverage (see Fig.~\ref{fig:footprint-cutsky}). We use the same $k_\mathrm{min}=0.02$ for all tracers in this section. We use the \textit{complete} EZmock to demonstrate the extra information gain from multitracer analysis. The covariance matrices are rescaled by 20 to match the future DESI comoving survey volume for overlapping LRG and ELG samples between redshift range $0.6<z<1.0$ \citep{DESI:2016fyo}\footnote{Our forecast here is conservative, since DESI has higher galaxy number densities than eBOSS and the multitracer method is more efficient with lower shot noise \citep{Seljak:2008xr,McDonald:2008sh}. Additionally, a simple rescaling does not account for the extra information coming from the wider survey window.}. The cut-sky window function multipoles and the correlation matrices are shown in Fig.~\ref{fig:window-cutsky} and Fig.~\ref{fig:corr_cutsky} respectively. Unlike the full-sky correlation matrices displayed in Fig.~\ref{fig:corr_full}, the cross correlation blocks of $P^L\text{-}P^X$ and $P^E\text{-}P^X$ show symmetric patterns. The correlation between the two auto power spectra at large scales is also clearer. In this section, we use the following data vector
\begin{equation}
    \bs{D} = \{ P_0^L, P_2^L, P_4^L, P_0^E, P_2^E, P_4^E, P_0^X, P_2^X, P_4^X \}
\end{equation}
to mimic the situation when angular systematics are well understood and cosmological information can be robustly extracted up to $\ell=4$.

\begin{figure*}
    \includegraphics[width=0.95\linewidth]{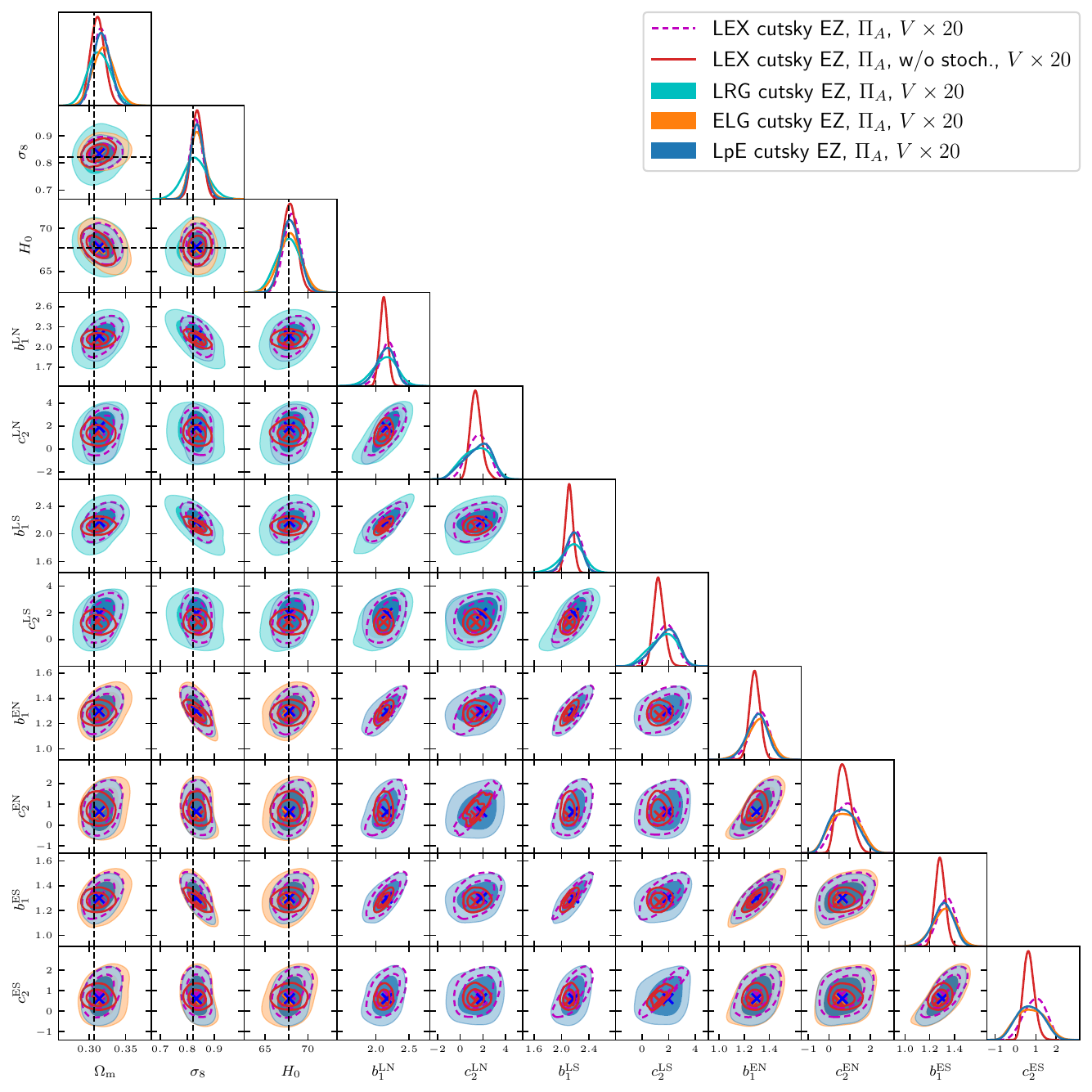}
    \caption{Full posterior plot for all cosmological parameters and EFT parameters derived from the cutsky \textit{complete} EZmocks. LN, LS, EN, ES represent LRG NGC, LRG SGC, ELG NGC, ELG SGC respectively.}
    \label{fig:cutsky-EZmock-contour-full}
\end{figure*}

\begin{figure*}
    \includegraphics[width=0.95\linewidth]{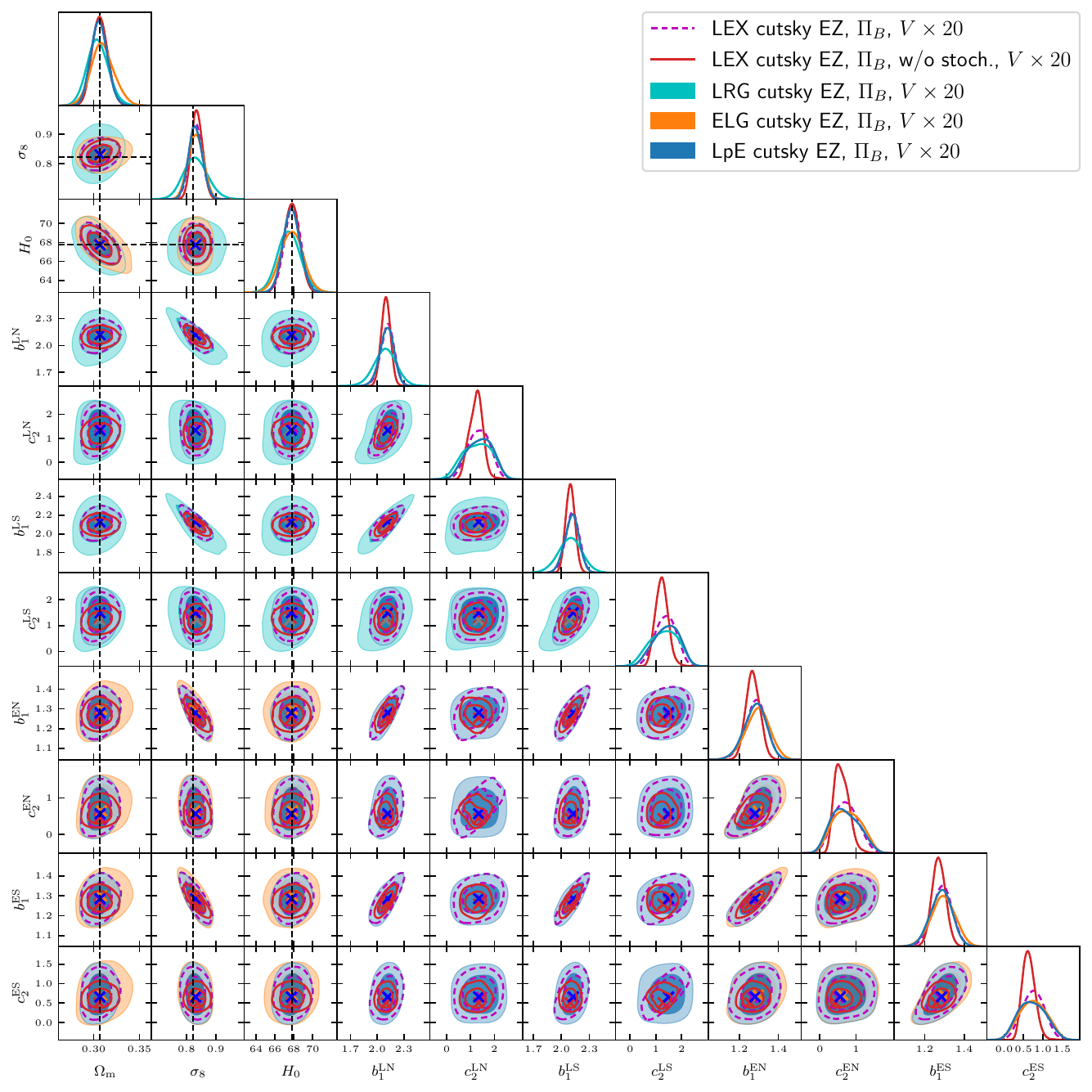}
    \caption{Same as Fig.~\ref{fig:cutsky-EZmock-contour-full}, but including a Gaussian prior on nuisance parameters to keep the theory perturbative (see Table.~\ref{tab:nuisance-prior}).}
    \label{fig:cutsky-EZmock-contour-full-perturbative}
\end{figure*}

The contour plot for cosmological parameters using the uninformative $\Pi_A$ prior are shown in Fig.~\ref{fig:cutsky-EZmock-contour}, and the relative improvements compared to using ELG data alone are summarized in Table~\ref{tab:EZmock-cutsky-improvement}. The full parameter posteriors are displayed in Fig.~\ref{fig:cutsky-EZmock-contour-full} and Fig.~\ref{fig:cutsky-EZmock-contour-full-perturbative} for the $\Pi_A$ prior and the $\Pi_B$ prior respectively. These plots also include a case combining the two auto power spectra\footnote{This is also the multitracer analysis, as the two auto power spectra follow the same structures \citep[see][]{Blake:2013nif}.}, allowing us to investigate the cosmological information in the cross power spectrum.

\begin{table}
    \centering
    \caption{The relative improvement on cosmological constraints (evaluated by comparing 1D $68$ per cent credible intervals) compared to using ELG power spectrum alone with different priors and configurations. $\mathcal{M}_A$ is the base model discussed in section~\ref{subsec:parameter-estimation} and $\mathcal{M}_B$ further fixes the stochastic terms in cross power spectrum to zero based on $\mathcal{M}_A$. Those results are derived from the cutsky \textit{complete} EZmock analysis with covariance matrices rescaled by 20.}
    \begin{tabular}{lccc}
    \hline
    & $\Omega_m$ & $\sigma_8$ & $H_0$ \\
    \hline\hline
    LpE+$\Pi_A$+$\mathcal{M}_A$ & $19.3\%$ & $9.6\%$ & $21.9\%$ \\
    \hline
    LEX+$\Pi_A$+$\mathcal{M}_A$  & $24.3\%$ & $13.5\%$ & $27.0\%$ \\
    \hline
    LEX+$\Pi_A$+$\mathcal{M}_B$  & $34.0\%$ & $24.9\%$ & $34.3\%$ \\
    \hline
    LpE+$\Pi_B$+$\mathcal{M}_A$  & $27.6\%$ & 9.6\% & $24.8\%$ \\
    \hline
    LEX+$\Pi_B$+$\mathcal{M}_A$  & $29.7\%$ & $14.3\%$ & $28.5\%$ \\
    \hline
    LEX+$\Pi_B$+$\mathcal{M}_B$  & $31.2\%$ & $26.6\%$ & $30.6\%$ \\
    \hline
    \end{tabular}
    \label{tab:EZmock-cutsky-improvement}
\end{table}

After the sky cuts, the cosmological constraints from two different tracers become comparable. The ELG sample yields better constraint on $\sigma_8$ due to it's smaller linear galaxy bias. By combining LRG and ELG, the constraints on $\Omega_m$ and $H_0$ are improved by $\sim 25\%$, and $\sigma_8$ by $\sim 15\%$, regardless of the prior used. If we fix the stochastic terms in the cross power spectrum to zero, improvements are further boosted: $\sim 25\%$ for $\sigma_8$. The full posterior plot shows that combining two auto power spectra improves the constraint on the linear bias parameter $b_1$ and thus breaks its degeneracy with cosmological parameters. The extra information provided by the cross power spectrum is marginal unless the stochastic terms are fixed. This is because the improvement is mainly absorbed by the higher order bias parameter $c_2$, which is less degenerate with cosmological parameters. However, fixing stochastic terms in the cross power spectrum significantly tightens the constraints on all parameters because in this case the cross power spectrum effectively breaks the degeneracy between stochastic terms and cosmological parameters in the auto power spectra. This improvement cannot be interpreted as the cosmic variance cancellation. Therefore, it is essential to investigate the relevance of this term in future studies with the aid of N-body based mocks and state-of-the-art multitracer galaxy population models. Fig.~\ref{fig:cutsky-EZmock-bar} summarizes the multitracer improvement at different $k_\mathrm{max}$ with different prior choices and configurations. Given the same setting, we find the improvement relative to the single-tracer analysis remains stable with different $k_\mathrm{max}$ cuts.


\section{Full parameter posteriors}
\label{appendix:b}
\begin{figure*}
    \includegraphics[width=0.95\linewidth]{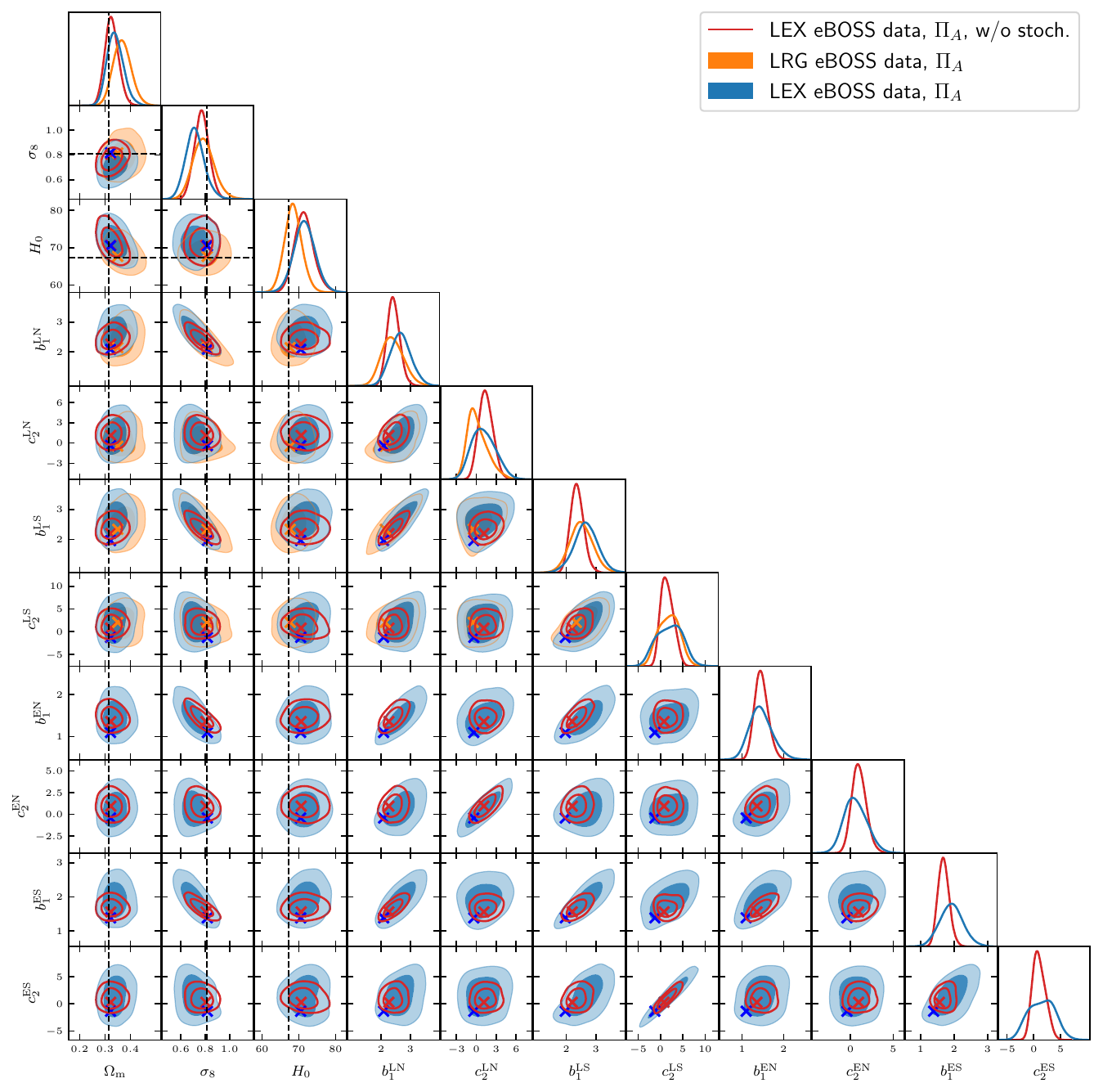}
    \caption{Full posterior plot for all cosmological parameters and EFT parameters derived from eBOSS data. LN, LS, EN, ES represent LRG NGC, LRG SGC, ELG NGC, ELG SGC respectively. For comparison, the black dashed line shows the best-fit of Planck18 baseline analysis.}
    \label{fig:data-contour-full}
\end{figure*}
\begin{figure*}
    \includegraphics[width=0.95\linewidth]{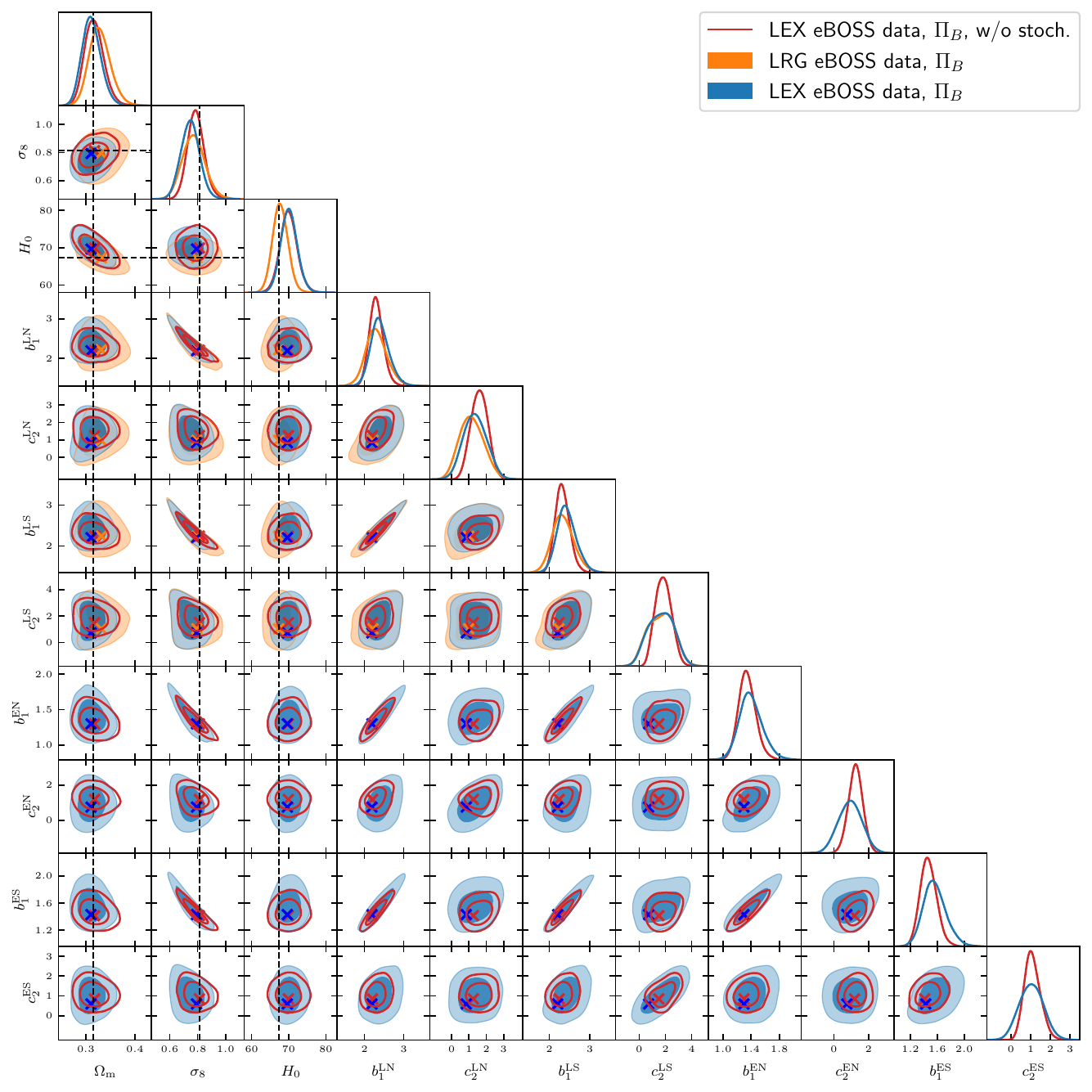}
    \caption{Same as Fig.~\ref{fig:data-contour-full}, but including a Gaussian prior on nuisance parameters to keep the theory perturbative (see Table.~\ref{tab:nuisance-prior}).}
    \label{fig:data-contour-full-perturbative}
\end{figure*}



\bsp	
\label{lastpage}
\end{document}